%% file: yunlong-icsme-2021.tex
\pdfoutput=1
\documentclass[conference]{IEEEtran}
\IEEEoverridecommandlockouts
\usepackage{amsmath,amssymb,amsfonts}
\usepackage{algorithmic}
\usepackage{graphicx}
\usepackage{textcomp}
\usepackage{xcolor}
\def\BibTeX{{\rm B\kern-.05em{\sc i\kern-.025em b}\kern-.08em
    T\kern-.1667em\lower.7ex\hbox{E}\kern-.125emX}}

\usepackage{booktabs}   
\usepackage{enumitem}
\usepackage{xspace}
\usepackage{multirow}
\usepackage{amsthm}
\usepackage{makecell}
\usepackage{subfig}
\usepackage[algoruled,boxed,vlined]{algorithm2e}
\usepackage{color,soul}
\usepackage{bbding}

\usepackage{amssymb}
\usepackage{pifont}
\newcommand{\cmark}{\ding{51}}%
\newcommand{\xmark}{\ding{55}}%
\usepackage{comment}
\usepackage{url}

\setul{0.5ex}{0.3ex}
    \definecolor{Red}{rgb}{1,0.0,0.0}
    \setulcolor{Red}

\newcommand{\tool}{FAPR\xspace}
\newcommand{\codefont}[1]{\footnotesize{\texttt{#1}}\normalsize}

\newcommand{\totalc}{20,897\xspace}
\newcommand{\ic}{6,676\xspace}

\newcommand{\totalcpp}{25,000\xspace}

\newcolumntype{L}[1]{>{\raggedright\let\newline\\\arraybackslash\hspace{0pt}}m{#1}}
\newcolumntype{C}[1]{>{\centering\let\newline\\\arraybackslash\hspace{0pt}}m{#1}}
\newcolumntype{R}[1]{>{\raggedleft\let\newline\\\arraybackslash\hspace{0pt}}m{#1}}
\newcolumntype{?}{!{\vrule width 1pt}}

\theoremstyle{definition}

\begin{document}

\title{\tool: Fast and Accurate Program Repair for Introductory Programming Courses}

\author{\IEEEauthorblockN{Yunlong Lu}
\IEEEauthorblockA{
\textit{Peking University}\\
luyunlong@pku.edu.cn}
\and
\IEEEauthorblockN{Na Meng}
\IEEEauthorblockA{
\textit{Virginia Tech}\\
nm8247@cs.vt.edu}
\and
\IEEEauthorblockN{Wenxin Li}
\IEEEauthorblockA{
\textit{Peking University}\\
lwx@pku.edu.cn}
}

\maketitle

\input{abstract}

\begin{IEEEkeywords}
Programming education, MOOC, 
program encoding, 
dynamic programming, 
program differencing, 
repair validation, 
program repair
\end{IEEEkeywords}

\vspace{-.5em}
\input{intro}
\vspace{-.5em}
\input{example}
\vspace{-.5em}
\input{approach}
\vspace{-.5em}
\input{implement}
\vspace{-.5em}
\input{evaluation}
\vspace{-.5em}
\input{threats}
\vspace{-.5em}
\input{related}
\vspace{-.5em}
\input{conclusion}

\bibliography{yunlong}

\end{document}


\title{Appendix for \tool: Fast and Accurate Program Repair for Introductory Programming Exercises}

\maketitle


\section{The Five Programming Problems Selected for Evaluation} \label{app:A}
\subsection{Problem 1678}
\paragraph{Description}
Given $k$ integers ($1<k<100$), each of which is no less than 1 and no more than 10, you are asked to write a program to count the occurrences of 1, 5 and 10 respectively.
\paragraph{Input}
There are two lines. The first line contains an integer $k$. The second line contains $k$ integers split by a single space.
\paragraph{Output}
Three lines containing the number of occurrences of 1, 5 and 10 respectively.
\paragraph{Sample Input} \mbox{}\\
5\\
\noindent 1 5 8 10 5
\paragraph{Sample Output} \mbox{}\\
1\\
\noindent 2\\
\noindent 1
\subsection{Problem 1689}
\paragraph{Description}
Show an array of values in reverse order. For example, the original array is 8, 6, 5, 4, 1. You are asked to change it to 1, 4, 5, 6, 8.
\paragraph{Input}
There are two lines. The first line contains an integer $n$ ($1<n<100$). The second line contains $n$ integers split by a single space.
\paragraph{Output}
One line of integers of the reversed array, split by a single space.
\paragraph{Sample Input} \mbox{}\\
5\\
\noindent 8 6 5 4 1
\paragraph{Sample Output} \mbox{}\\
1 4 5 6 8
\subsection{Problem 1703}
\paragraph{Description}
A positive integer is said to be related to 7 if it is divisible by 7, or one of the digit in its decimal representation is 7. Calculate the sum of the squares of all positive integers less than or equal to $n$ ($n<100$) that is not related to 7.
\paragraph{Input}
One line with a single integer n.
\paragraph{Output}
One line with an integer, which is the sum required.
\paragraph{Sample Input} \mbox{}\\
21
\paragraph{Sample Output} \mbox{}\\
2336
\subsection{Problem 1716}
\paragraph{Description}
There are several students in the class. Given the age of each student in integers, find the average age of all the students in the class, rounded to two decimal places.
\paragraph{Input}
The first line contains an integer $n$ ($1<n<100$), which is the number of students. Each of the next $n$ lines contains an integer ranging from 15 to 25, which is the age of a student.
\paragraph{Output}
One line with a floating-point number rounded to two decimal places, which is the average age required.
\paragraph{Sample Input} \mbox{}\\
2\\
\noindent 18\\
\noindent 17
\paragraph{Sample Output} \mbox{}\\
17.50
\subsection{Problem 1720}
\paragraph{Description}
Jingjing's friend Beibei wants to invite Jingjing to go to the exhibition next week, but Jingjing has to attend classes on Monday, Wednesday and Friday. Help Jingjing decide whether she can accept Beibei's invitation, Print ``YES'' if she can and ``NO'' if she cannot.
\paragraph{Input}
One line with an integer representing the day when Beibei invited Jingjing to see the exhibition, 1~7 for Monday to Sunday.
\paragraph{Output}
One line containing ``YES'' if Jingjing can accept Beibei's invitation and ``NO'' if she cannot. Note that both ``YES'' and ``NO'' are in capital letters!
\paragraph{Sample Input} \mbox{}\\
2
\paragraph{Sample Output} \mbox{}\\
YES

\section{More code examples}
Here we demonstrate five programs and the generated repairs from \tool, one for each problem. The repair relevance (RR) and repair usefulness (RU) of each repair from user study are given as well as the explanation of these repairs. These programs are selected to show how repairs of different RR look like.

\subsection{Example from Problem 1678}\mbox{}\\
\begin{figure}[h]
\includegraphics[width=\linewidth]{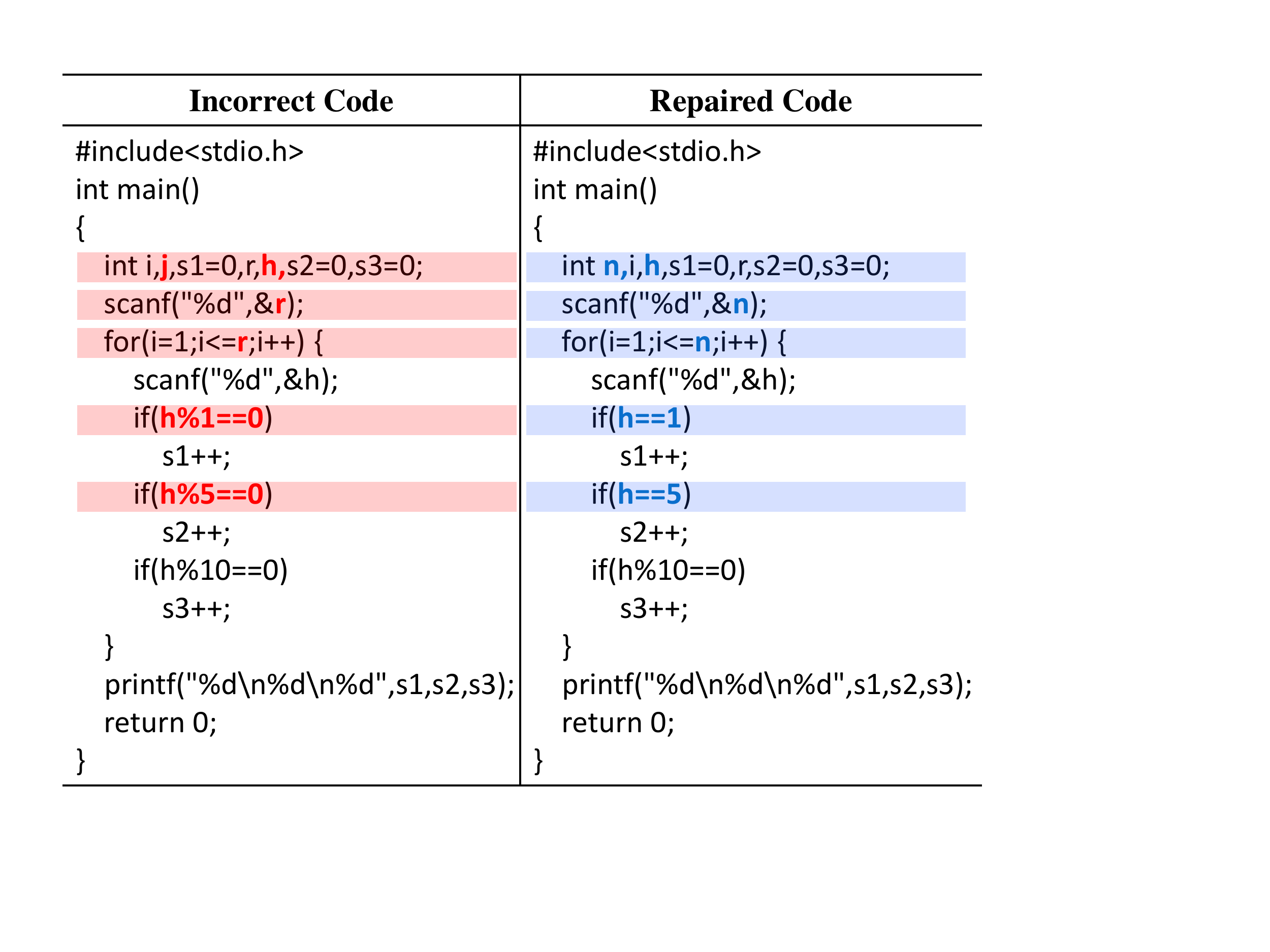}
\caption{An incorrect program from problem 1678 and the generated repair, with RR = 3 and RU = 5}
\label{figexample1}
\end{figure}

\paragraph{Explanation for error}
In Problem 1678, the program should count the number of 1, 5 and 10 among the integers from the input. However, the incorrect solution in Figure~\ref{figexample1} uses the modulo operator to check the divisibility of each integer for 1, 5 and 10, instead of equality, probably because of a misunderstanding of the original problem.

\paragraph{Explanation for repair}
The most natural way to fix this incorrect program is to change the conditions of three if-statement from divisibility check to equality check. However, since the problem ensures that each number from input is no less than 1 and no more than 10, the last divisibility check can remain unchanged because 10 is the only number to be divisible by 10 in the given range. So the correct and smallest repair should only change the first two conditions.

The repair generated by \tool for this solution gets a RR score of 3, because it is correct and fixes the bug, but includes unnecessary changes that move the definition of variable $h$ and use a different variable $n$ for the number of input. It get a RU score of 5 i.e. is considered very useful by the volunteer, probably because the volunteer thinks that the error in original solution is fixed exactly, though some minor changes are also included.

\subsection{Example from Problem 1689}\mbox{}\\
\begin{figure}[h]
\includegraphics[width=0.7\linewidth]{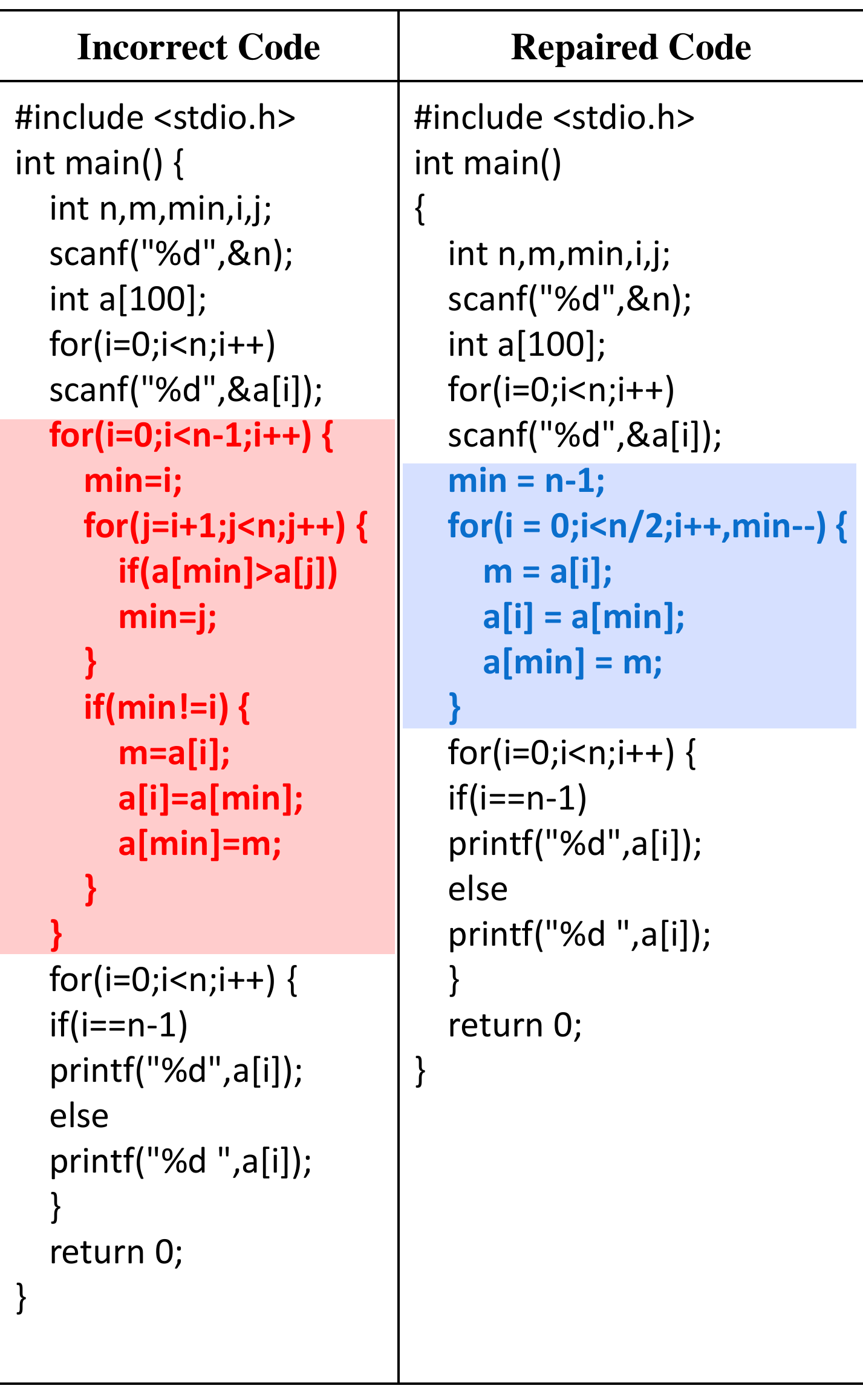}
\caption{An incorrect program from problem 1689 and the generated repair, with RR = 2 and RU = 5}
\label{figexample2}
\end{figure}

\paragraph{Explanation for error}
In Problem 1689, the program should reverse a given array. However, the incorrect solution in Figure~\ref{figexample2} implements the algorithm of insertion sorting and sorts the given array in ascending order, probably because the student misunderstood the problem when he/she sees the sample output (1 4 5 6 8) to be in ascending order, which is a coincidence.

\paragraph{Explanation for repair}
To fix this incorrect program, \tool replaces a large chunk of code from the original solution. Namely, it replaces the part of insertion sorting with a for-loop that correctly reverses the array. It is worth mention that the generated for-loop uses the same variables in the original solution. Instead of $i$ and $j$ as common sense, it uses $i$ and $min$ as array index to avoid unnecessary code change because the original solution uses them, as an evidence of the strong adaptation of \tool.

The RR score of this repair is only 2, because the generated repair significantly changes the original algorithm design. However, such low repair relevance is unavoidable for any correct repair because the algorithm design of original solution i.e. insertion sorting is wrong. The RU score of this repair is also marked as 5 i.e. very useful by our volunteer, because the generated repair correctly fixes the algorithm used.

\subsection{Example from Problem 1703}\mbox{}\\
\begin{figure}[h]
\includegraphics[width=0.8\linewidth]{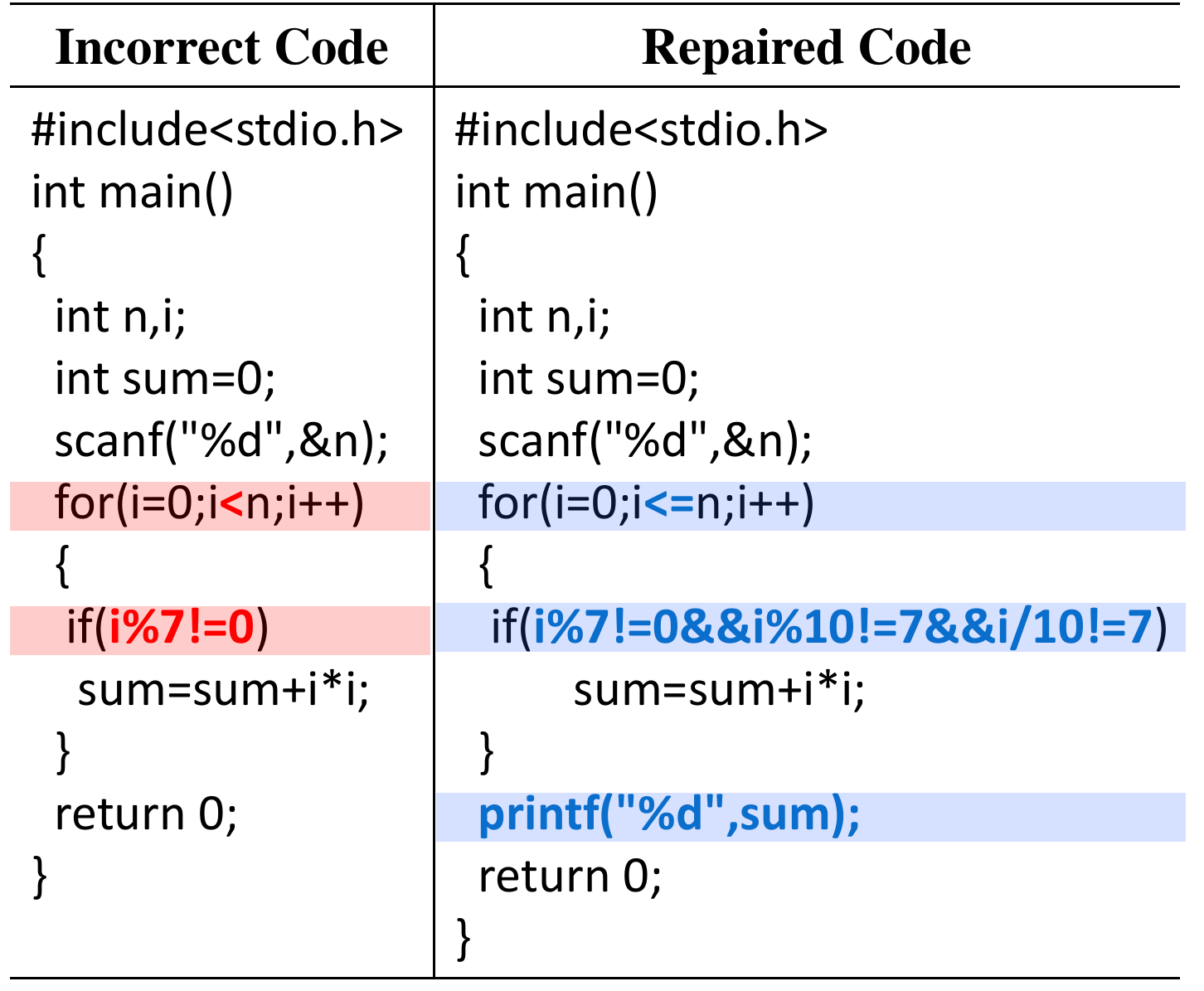}
\caption{An incorrect program from problem 1703 and the generated repair, with RR = 4 and RU = 5}
\label{figexample3}
\end{figure}

\paragraph{Explanation for error}
In Problem 1703, the program should enumerate each number no more than $n$ and check if it is related to 7 by checking its divisibility by 7 and each digit. The incorrect solution in Figure~\ref{figexample3} makes three mistakes. First, it only enumerates each number under $n$, missing the consideration of $n$ itself. Second, it only checks the divisibility of each number by 7, but a number is related to 7 can also have 7 in one of its digits according to the problem, which is not considered in this program. Third, it forgets to print the result to standard output.

\paragraph{Explanation for repair}
The repair generated by \tool consists of three edits, each of which exactly fixes one mistake of the original solution. Note that $n$ is smaller than 100 in the problem description, so checking each digit only needs to check the ones place and tens place of a number, instead of a more complex solution to use a loop. The RR score of this repair is 4 because it correctly fixes the bug without introducing unnecessary code changes. The RU score of this repair is marked as 5 i.e. very useful by our volunteer.

\subsection{Example from Problem 1716}\mbox{}\\
\begin{figure}[h]
\includegraphics[width=\linewidth]{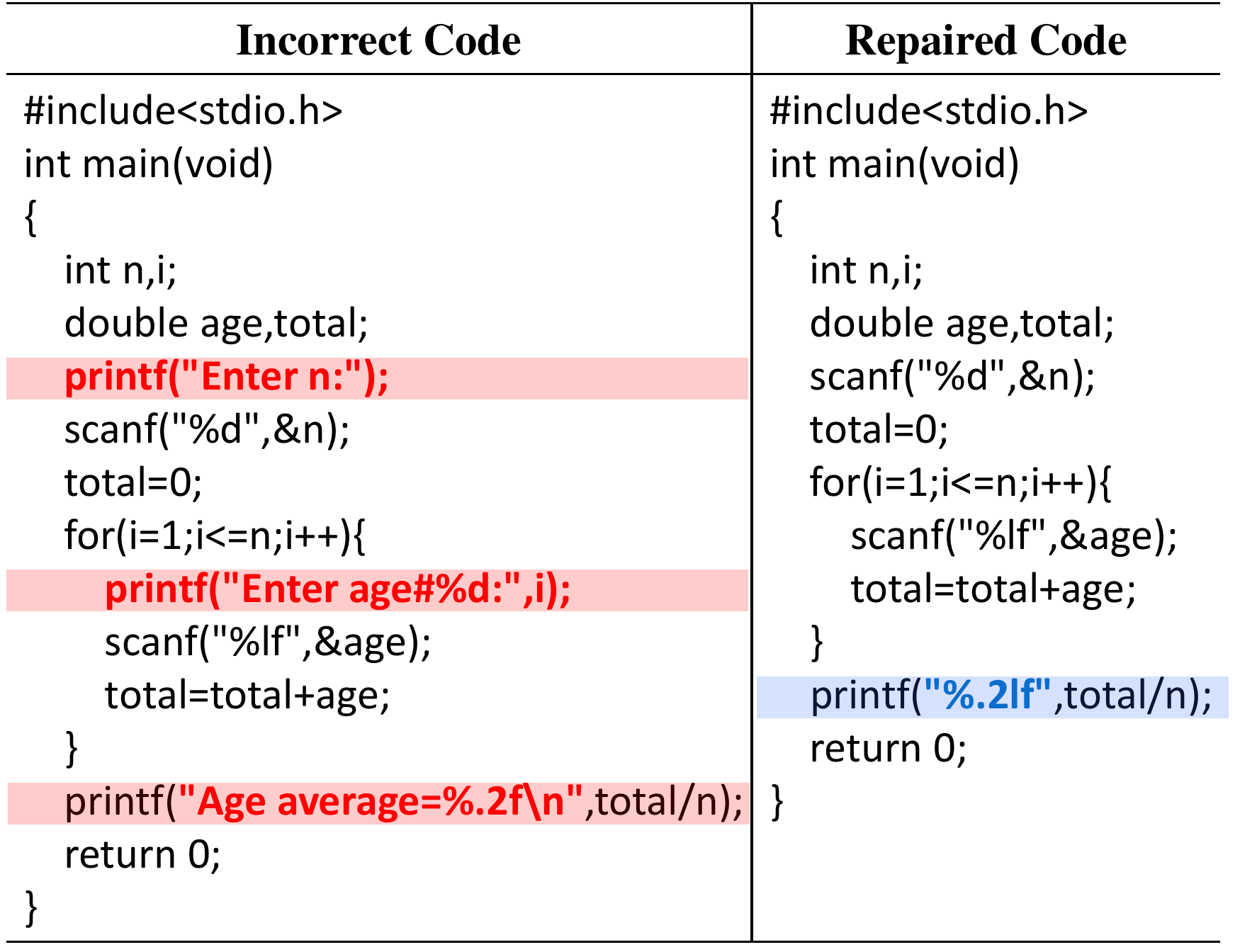}
\caption{An incorrect program from problem 1716 and the generated repair, with RR = 4 and RU = 5}
\label{figexample4}
\end{figure}

\paragraph{Explanation for error}
In Problem 1716, the program should read a list of numbers and calculate their mean. The incorrect program in Figure~\ref{figexample4} makes a common mistake by beginners to output extra prompts in the program as the textbook does, which is not needed and fails tests in an online judge.

\paragraph{Explanation for repair}
The repair generated by \tool fixes the mistake exactly, by removing the extra prompts before input and output of the original program. The RR score of this repair is 4 because it correctly fixes the bug without introducing unnecessary code changes. The RU score of this repair is marked as 5 i.e. very useful by our volunteer.

\subsection{Example from Problem 1720}\mbox{}\\
\begin{figure}[h]
\includegraphics[width=\linewidth]{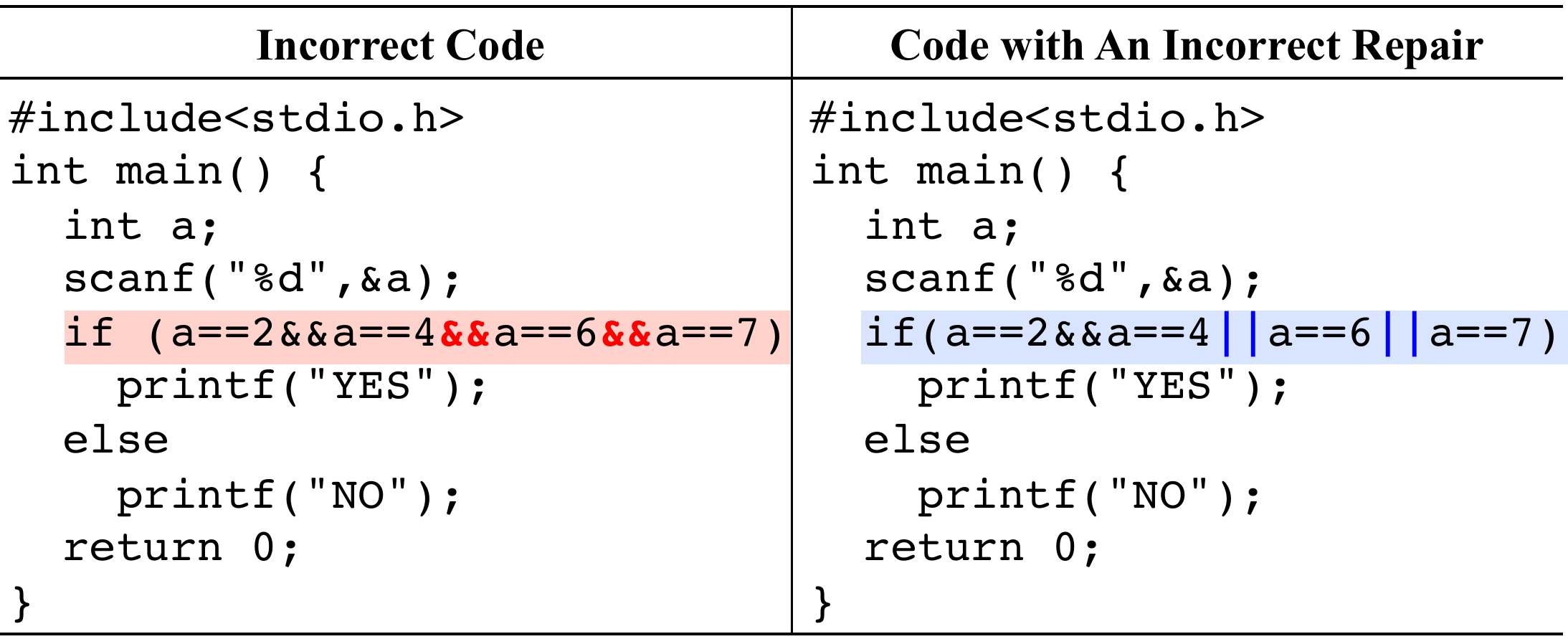}
\caption{An incorrect program from problem 1720 and the generated repair, with RR = 1 and RU = 4}
\label{figexample5}
\end{figure}

\paragraph{Explanation for error}
In Problem 1720, the program should make a simple judgement, and output ``YES'' if the input is 1, 3 or 5, and ``NO'' if it is 2, 4, 6 or 7. The incorrect solution in Figure~\ref{figexample5} mistakenly concatenates multiple conditions by operator and(\&\&) instead of operator or(||), causing the program to output ``NO'' whatever the input is.

\paragraph{Explanation for repair}
The repair generated by \tool only modifies two of the three operator and(\&\&), and the RR score of it is marked as 1 i.e. the repair is incorrect by our volunteer. We further analyze the reason of it and find something interesting. The test cases of this problem only cover five situations i.e. the input is 1, 3, 5, 6 and 7. \tool first generates repair consisting of three edits, but drop the first one when minimizing repair set because dropping this edit also passes all test cases. This example is one that shows the common limitation of testing-based APR techniques. However, from another point of view it also shows the effectiveness of our tool to generate a minimal repair that passes test. The RU score of this repair is marked as 4 i.e. this repair is somewhat useful by our volunteer, because it points out the exact mistakes of the original program and is inspiring, though does not fully fix it.

%% file: abstract.tex
\begin{abstract}
In introductory programming courses, it is challenging for instructors to provide debugging feedback on students' incorrect programs. 
Some recent tools automatically offer program repair feedback by identifying any differences between incorrect and correct programs, but suffer from issues related to scalability, accuracy, and cross-language portability. This paper presents \tool---our novel approach that suggests repairs based on program differences in a fast and accurate manner.  
\tool is different from current tools in three aspects. 
First, it encodes syntactic information into token sequences to enable high-speed comparison between incorrect and correct programs. Second, to accurately extract program differences, \tool adopts a novel matching algorithm that maximizes 
token-level matches and minimizes statement-level differences. 
Third, \tool relies on testing instead of static/dynamic analysis to validate and refine candidate repairs, so it eliminates the language dependency or high runtime overhead incurred by complex program analysis.

We implemented \tool to suggest repairs for both C and C++ programs; our experience shows the great cross-language portability of \tool. More importantly, we empirically compared \tool with a state-of-the-art tool Clara. 
\tool suggested repairs for over 95.5\% of incorrect solutions. We sampled 250 repairs among \tool's suggestions, and found 89.6\% of the samples to be minimal and correct. \tool outperformed Clara by suggesting repairs for more cases, creating smaller repairs, producing higher-quality fixes, and causing lower runtime overheads. 
Our results imply that \tool can potentially help instructors or TAs to effectively locate bugs in incorrect code, and to provide debugging hints/guidelines based on those generated repairs.
\end{abstract}

%% file: intro.tex
\section{Introduction}

In introductory programming courses, students may commit various mistakes when they learn to program. It is very challenging or even impossible for instructors and teaching assistants (TAs) to provide one-on-one programming instructions, and to help all students debug their incorrect code. 
This challenge has become even more severe and intractable during the COVID-19 pandemic, as Massive Online Open Courses (MOOC) see exponential growth in enrollments and thousands of students submit coding solutions online~\cite{mooc}. 
While lots of students desperately need teachers' suggestions on code debugging, instructors and TAs usually have limited time availability and programming capabilities. Both teachers and students need tools to automatically provide debugging feedback or program repair suggestions. 
 
Some recent tools were built to suggest program repairs by comparing  correct and incorrect solutions to the same coding problem~\cite{Singh13,pu2016skp,Kaleeswaran16,Gulwani18,Perry19}. 
For instance, 
Clara~\cite{Gulwani18} suggests edits by (1) clustering correct solutions based on the equivalence of program execution status, and (2) comparing any given incorrect code with the representative programs of each cluster to find a minimum repair. However, the high runtime overhead of Clara's dynamic analysis can make the tool pretty slow, when lots of students submit incorrect code simultaneously to seek for debugging feedback. The  heavy reliance on sophisticated dynamic analysis also compromises Clara's applicability to code written in different programming languages. 
Sk\_p~\cite{pu2016skp} trains a sequence-to-sequence neural network using token sequences of correct programs, and leverages the trained model to suggest repairs for given incorrect code. However, the suggested repairs are usually incorrect, achieving 13\%--49\% accuracy. 

To better help teachers provide debugging feedback on students' solutions to introductory programming problems, 
we need to tackle two challenges:
\begin{itemize}
\item[1.] Lightweight program comparison methods (e.g., token-level code comparison) are usually fast but may suggest inaccurate repairs, while heavyweight alternatives (e.g., comparing execution traces) are slow but may suggest more accurate repairs. We need to better balance speed with accuracy. 
\item[2.] As students may submit programs written in distinct programming languages, we need an approach that is generally applicable to different languaged programs. Thus, the less language-specific static or dynamic analysis is in use, the more portable a new approach can become. 
\end{itemize}
\begin{figure*}
\centering
\includegraphics[width=0.82\linewidth]{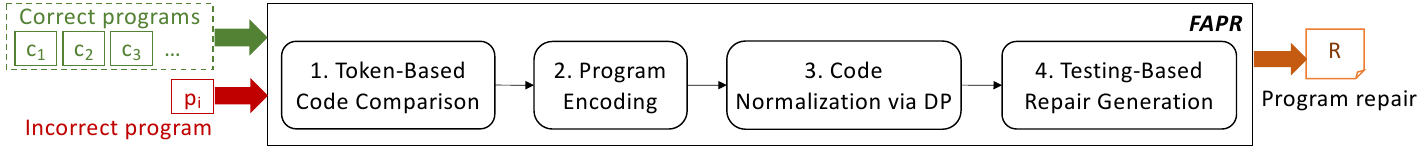}
\vspace{-.6em}
\caption{The overview of \tool}\label{fig:overview}
\vspace{-1.5em}
\end{figure*}
We designed and implemented \tool (short for ``\underline{F}ast and \underline{A}ccurate \underline{P}rogram \underline{R}epair'') to overcome both challenges.
\emph{We envision \tool to be used by teachers or automatic systems}. When teachers look at \tool's repair for any incorrect program, they can  quickly offer debugging hints/guidelines to students. When \tool is integrated into feedback generation systems, those systems can rely on the generated repairs to produce various kinds of feedback (e.g., bug location, high-level natural-language description on fixes, or partial edits for fixes). 
As shown in Fig.~\ref{fig:overview},  \tool  
has four phases. 
Given an incorrect program $p_i$, \tool first extracts the \textbf{longest common subsequence (LCS)} between the tokens of $p_i$ and all correct solutions, in order to quickly locate up to 100 programs $P$ that are most similar to $p_i$ as candidate references.  

The code differences captured by above-mentioned na\"ive token sequence comparison cannot be directly used to repair code for two reasons. First, some token differences (e.g., distinct variable usage) are irrelevant to software bugs and should not be suggested as repairs. Second, 
when multiple alternative LCSs exist between two programs, only the LCSs that minimize the number of edited statements are more desirable.  Thus, in Phase II, for each selected candidate reference $p_j\in P$, \tool revises its token sequence based on \textbf{the depth information of each token} (i.e., the number of AST nodes on the path from root to leaf). Based on such program encoding, in Phase III, 
\tool uses a matching algorithm 
to quickly select the best LCS that minimizes statement-level differences. 
\tool also uses the selected LCS to normalizes $p_j$ to $p_j'$ by replacing its distinct variable usage with that of $p_i$. 
In Phase IV, \tool explores subsets of the code differences between $p_j'$ and $p_i$ to tentatively repair $p_i$, and uses testing to validate each repair.

We did two experiments to evaluate \tool. In one experiment, we applied \tool to the \totalc C programs and \totalcpp C++ programs collected from an online judge system---OpenJudge~\cite{openjudge}. The collected data corresponds to five programming problems used in the course ``Introduction to Computing''. 
Among the \ic given incorrect C programs, \tool suggested repairs for 99\% of programs and the repaired programs pass all tests. In total, \tool spent only 306 seconds to preprocess the 14,221 correct C solutions; additionally, it spent 2.8 seconds on average to suggest each repair. 
We observed similar phenomena when applying \tool to C++ programs. 
In the other experiment, we randomly sampled 250 C repairs suggested by \tool,  
and recruited 5 students (assuming them to be TAs) to separately examine 50 samples. This manual evaluation shows that \tool's repairs have high quality: 98.4\% of them (246/250) are correct fixes, while 89.6\% of them (224/250) are minimum and correct.

To empirically compare \tool with prior work, we also applied Clara~\cite{Gulwani18} to the same C program dataset and repeated the above-mentioned two experiments. Our evaluation shows that \tool outperformed Clara by proposing more fixes with higher quality. 
More importantly, \tool has much lower runtime overheads. Clara spent in total 62 hours clustering correct C solutions before repairing any code; it spent 31.2 seconds 
 to suggest each repair.

In summary, our research contributions are as below:
\begin{itemize}
\item We developed a novel approach \tool to suggest program repairs in a fast and accurate manner. Different from prior work, \tool does not use any complex program analysis framework and is highly portable. Our experience with C/C++ programs also evidences \tool's portability. 
\item \tool encodes tree-level information into token sequences, and applies a dynamic programming (DP) algorithm to compare alternative LCSs. 
Our novel program encoding and DP algorithm
 enable \tool to balance speed with accuracy. 
\item We conducted a comprehensive evaluation to (1) assess \tool's applicability to C and C++ programs, (2) evaluate \tool's efficiency and the quality of its repairs, and (3) compare \tool with Clara. Our evaluation shows that \tool's repairs are usually correct, minimal, and useful. 

\end{itemize}

%% file: example.tex
\section{A Running Example}\label{sec:example}
This section overviews our approach with a running example drawn from the introductory CS course ``Introduction of Computing''. 
There is a coding exercise problem in the course: 

\begin{table}[h]
\footnotesize
\noindent\begin{tabular}{|p{8.4cm}|}
	\hline
	\textbf{Array Reversal:} Write a C/C++ program to reverse any given array. For example, given the array (1 4 5 6 8), the program outputs (8 6 5 4 1).	
	\\
	\hline
\end{tabular}
\end{table}

\begin{figure*}
\centering
\subfloat[Incorrect code $p_i$\label{fig:ic}]{
\includegraphics[width = 0.18\linewidth]{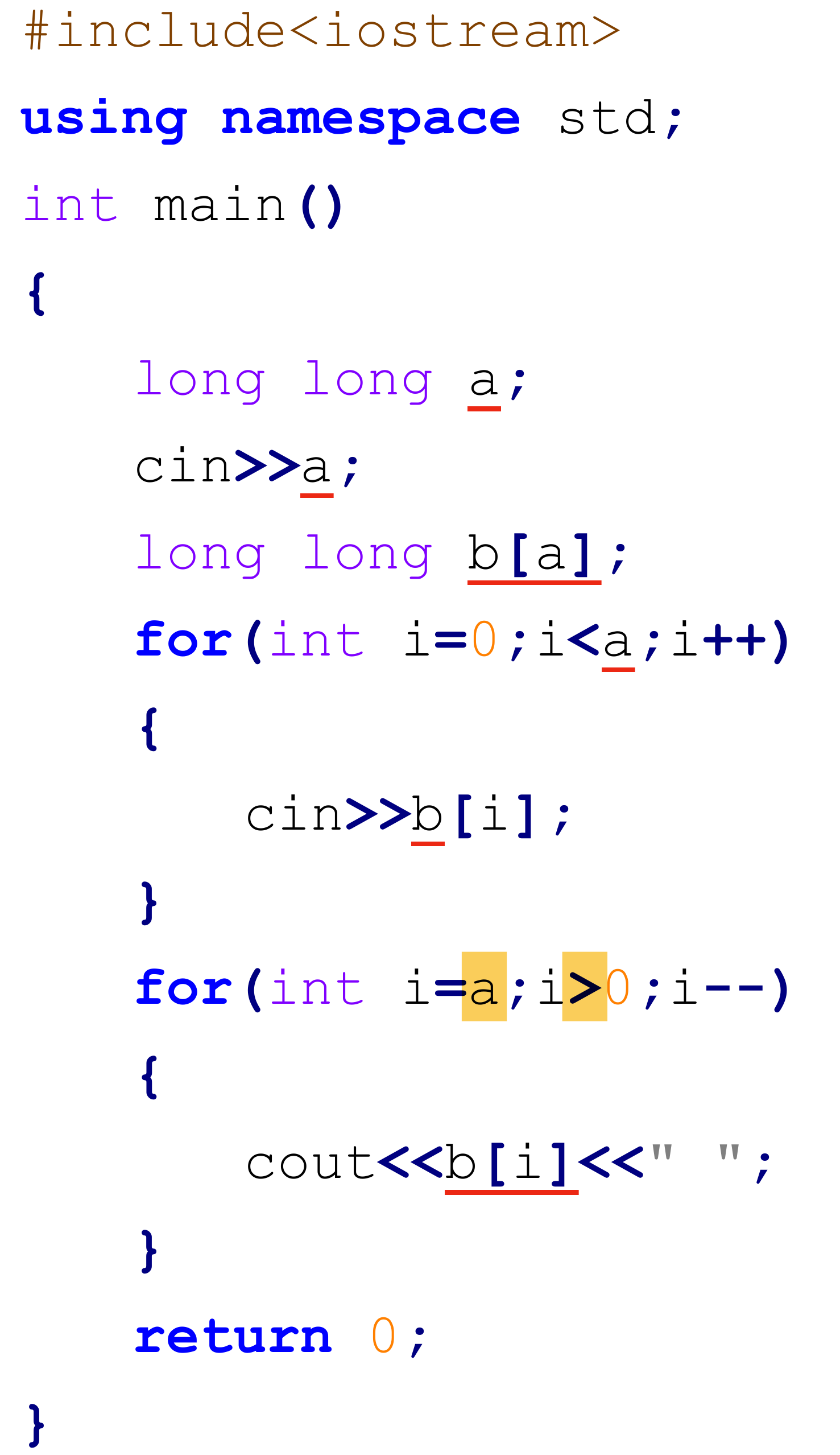}
}
\hspace{2em}
\subfloat[A correct solution $p_j$\label{fig:cc}]{
\includegraphics[width = 0.2\linewidth]{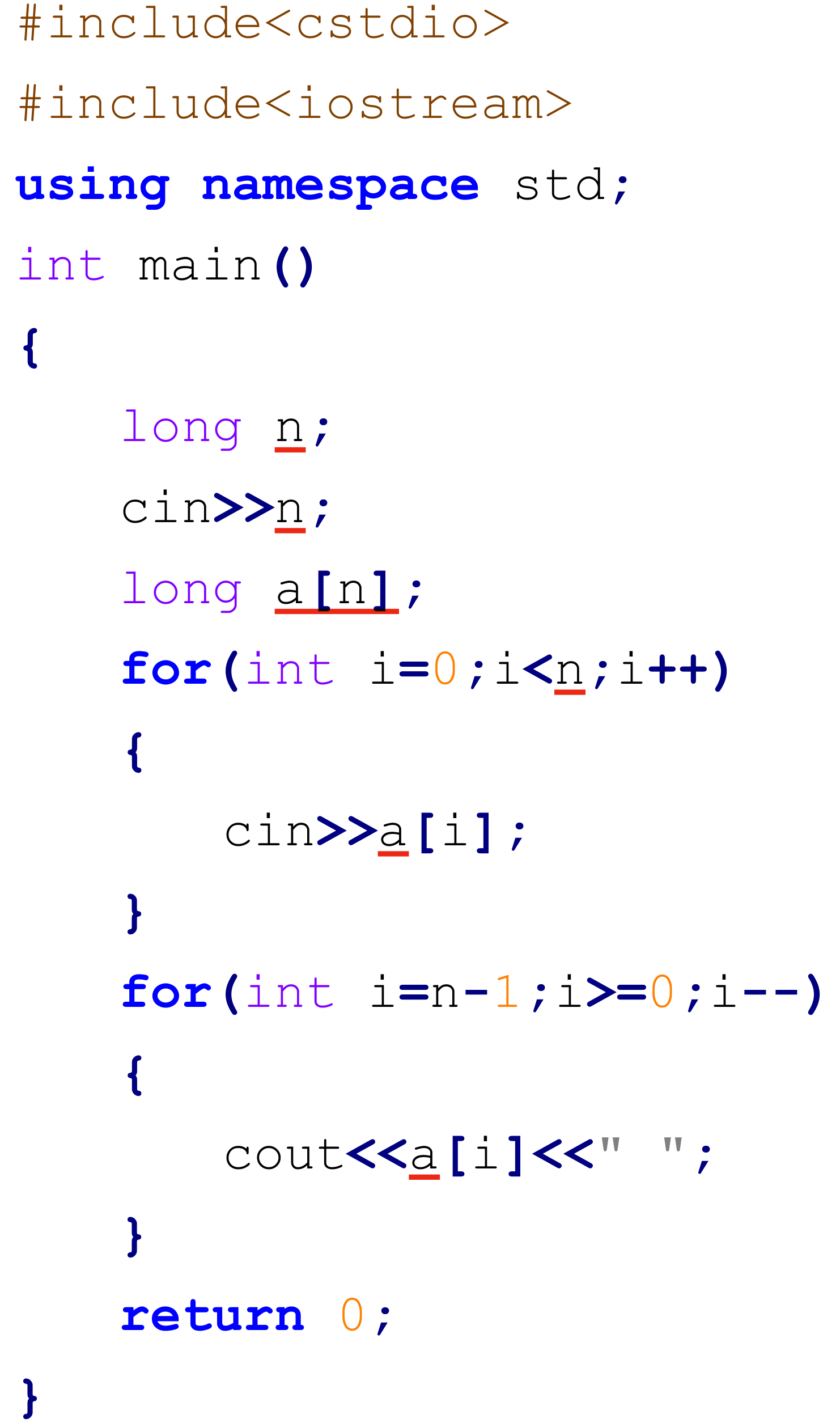}
}
\hspace{2em}
\subfloat[The normalized code $p_j'$ for $p_j$\label{fig:nc}]{
\includegraphics[width = 0.2\linewidth]{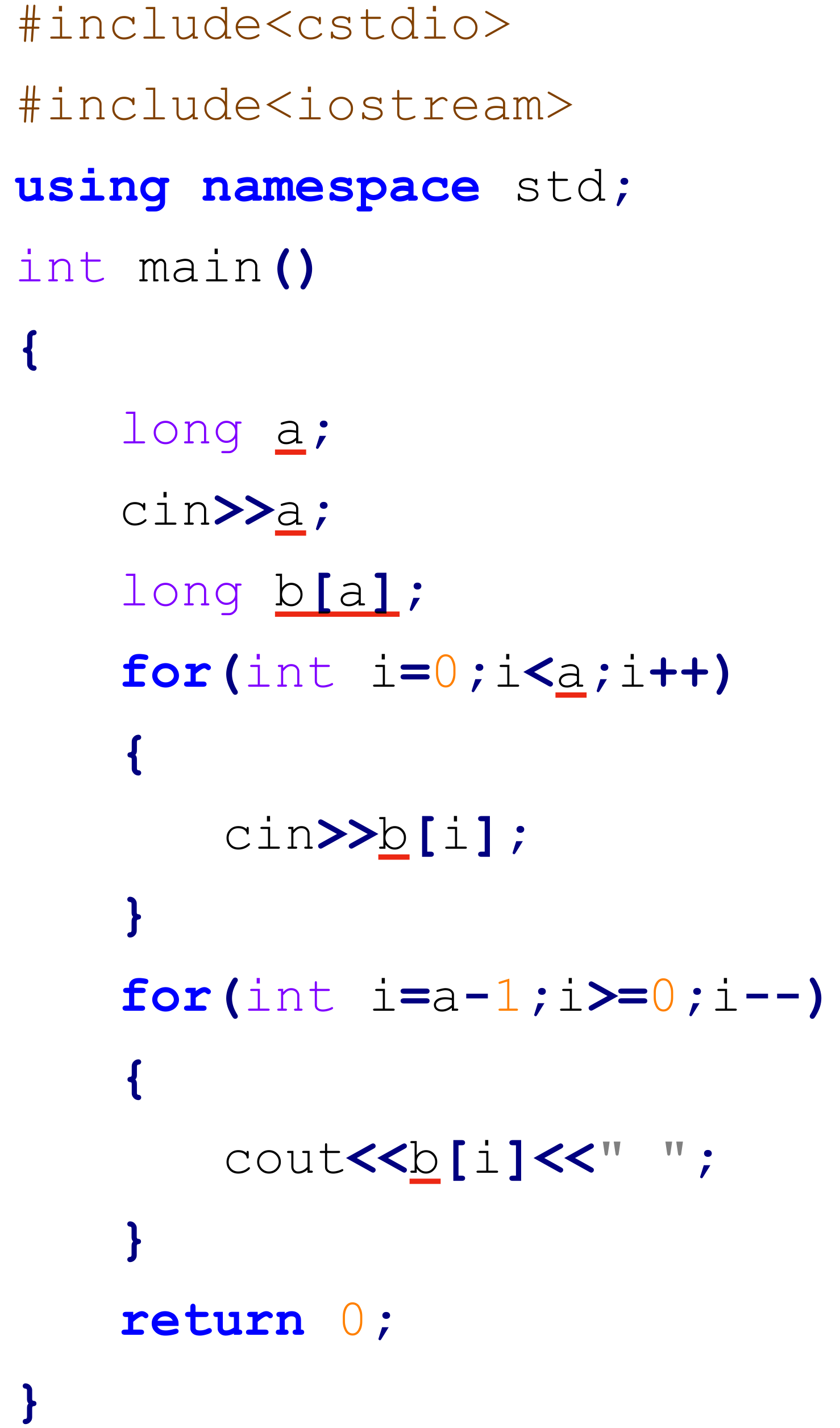}
}
\vspace{-.6em}
\caption{Demonstration of an incorrect program $p_i$, a correct program $p_j$,  and its normalized version $p_j'$}
\label{fig:example}
\vspace{-1.5em}
\end{figure*}

\noindent
Suppose that a student creates an incorrect program solution $p_i$ by enumerating array indexes in a wrong value range, as highlighted by \hl{yellow} in Fig.~\ref{fig:ic}. Namely, in the second \codefont{for}-loop, ``\codefont{int i=a; i>0; i$--$}'' should be ``\codefont{int i=a-1; i>=0; i$--$}''. When this student sends $p_i$ to an instructor Alex for debugging help, Alex can apply \tool to $p_i$ to quickly learn about the program repair before guiding that student to debug code. In particular, \tool consists of four phases. 

\paragraph*{Phase I. Code Comparison} \tool compares $p_i$ with existing correct solutions to retrieve programs $P$ most similar to $p_i$. Specifically for any two programs under comparison, \tool tokenizes the source code to extract an LCS $l$. Although $l$ is not directly used to compute repairs, \tool adopts it to quickly filter out solutions dissimilar to $p_i$ and to roughly rank the remaining ones based on similarities. 
We regard $P$ as \textbf{candidate references} usable to fix $p_i$. Fig.~\ref{fig:cc} shows a candidate reference selected for the incorrect solution.

\paragraph*{Phase II. Program Encoding} To facilitate more accurate program comparison between $p_i$ and $P$, \tool encodes tokens' relative location information on ASTs into token sequences. We refer to the resulting encoding \textbf{enhanced token sequences}. 

\paragraph*{Phase III. Code Normalization} Between the enhanced token sequences of two given programs $p_i$ and $p_j\in P$, \tool uses an efficient algorithm to explore all alternative LCSs 
 and select the best one with minimum statement-level differences. 
\tool also infers identifier mappings between those programs to normalize the code representation of $p_j$. 
For our example, $p_j$ is converted to $p_j'$ (see Fig.~\ref{fig:nc}), with its unique variable usage systematically replaced by that of $p_i$ (see the \ul{underlined variable usage} in Figures~\ref{fig:cc} and~\ref{fig:nc}).

\paragraph*{Phase IV. Repair Generation} \tool executes each normalized program $p_j'$ with all test cases to refine candidate references. In particular, if $p_j'$ passes all tests, the code differences between $p_j'$ and $p_i$ are regarded as \textbf{candidate repairing edits $\boldsymbol{R_j}$}. 
\tool then uses testing to iteratively explore and validate subsets of candidate edits. 
 This phase returns the minimized repair---a sequence of customized code edits---to Alex. 

%% file: approach.tex
\section{Approach}
This section explains individual phases with more details.

\vspace{-.5em}
\subsection{Code Comparison}\label{sec:phase1}

\tool first tokenizes code by taking two steps. In Step 1, it adopts an off-the-shelf tool srcML~\cite{srcml} to parse code, create Abstract Syntax Trees (AST), and store each AST into an XML file. In Step 2, \tool performs a depth-first traversal of the XML tree (i.e., AST) and collects all leaf nodes (except $\epsilon$-nodes) to build the token sequence. Notice that we intentionally derive token sequences from XML files instead of using a scanner because in later phases, \tool needs to revisit ASTs for code normalization and repair generation. 
 To eliminate the extra runtime overhead of (1) invoking a scanner and (2) repetitively parsing code, \tool parses each program only once and saves ASTs to XML files. 
When comparing two token sequences (e.g., $s_i$ and $s_j$), \tool extracts LCS by (1) matching variable and function names based on token type (i.e., identifier), and (2) matching other tokens based on content (i.e., lexemes~\cite{aho06}). 
The extracted LCS serves two purposes. First, it is used to compute similarity score as below:

\vspace{-1em}
\small
\begin{equation*}
Similarity (s_i, s_j) = \frac{length(LCS)}{Average (length(s_i), length(s_j))}.
\end{equation*} 
\normalsize 
Second, it will facilitate later comparison between program encodings (see Section~\ref{sec:phase2}). 
For our running example, the first \codefont{for}-loop headers of $p_i$ and $p_j$ have common tokens extracted as below (see underlined text):

\codefont{\underline{for} \underline{(} \underline{int} \underline{i} \underline{=} \underline{0} \underline{;} \underline{i} \underline{<} \underline{a} \underline{;} \underline{i} \underline{++} \underline{)} }

\codefont{\underline{for} \underline{(} \underline{int} \underline{i} \underline{=} \underline{0} \underline{;} \underline{i} \underline{<} \underline{n} \underline{;} \underline{i} \underline{++} \underline{)} }

\noindent
Literally, \tool considers all token pairs at corresponding locations to match. This is because the non-identifier tokens have matching content (e.g., \codefont{(for, for)}), and the identifiers have the same token type although \codefont{a} and \codefont{n} are different. \tool matches type names (i.e., \codefont{int}) by content instead of token type, because divergent type usage indicates bugs and/or fixes.

\begin{table}
\scriptsize
\centering
\vspace{.5em}
\begin{tabular}{|ll|}
\hline
$p_i$: & \codefont{\underline{\#} \underline{include} \underline{<iostream>}}\\
$p_j$'s alt1: & \codefont{\# include <cstdio> \underline{\#} \underline{include} \underline{<iostream>}}\\
$p_j$'s alt2: & \codefont{\underline{\#} include <cstdio> \# \underline{include} \underline{<iostream>}}\\
$p_j$'s alt3: & \codefont{\underline{\#} \underline{include} <cstdio> \# include \underline{<iostream>}}\\
\hline
\end{tabular}
\vspace{-0.6em}
\captionof{figure}{Three alternative LCSs \codefont{\#include}'s of $p_i$ and $p_j$}
\label{fig:alts}
\vspace{-1.5em}
\end{table}

\begin{figure}
\includegraphics[width=\linewidth]{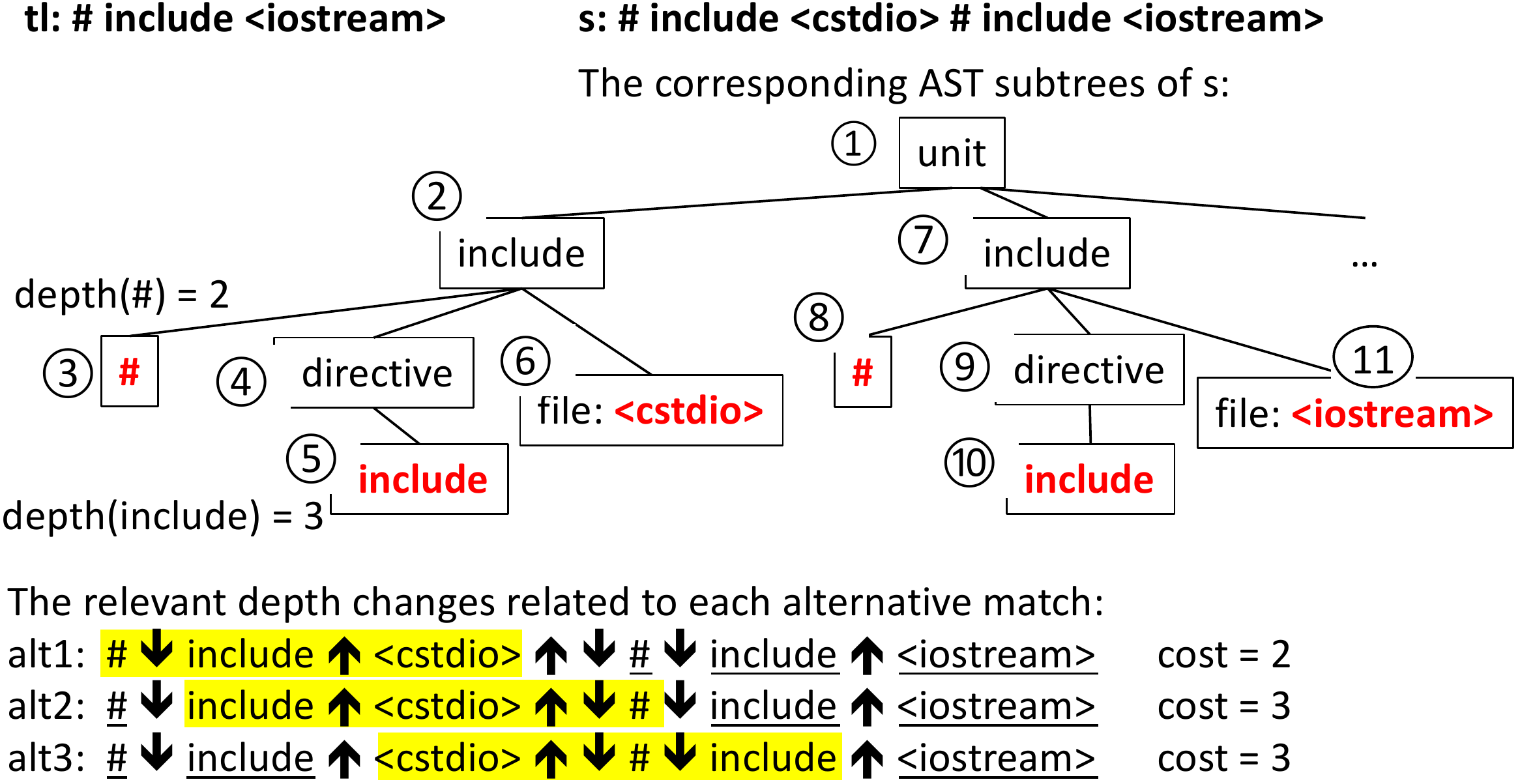}
\vspace{-1.6em}
\caption{The cost comparison between alternative LCSs}\label{fig:lcs-optimize}
\vspace{-1.5em}
\end{figure}

In particular, srcML is a multi-language parser that converts source code of C, C++, C\#, and Java to XML files. As \tool analyzes and compares code based on XML representations, it obtains the multi-language processing capability almost for free. To expand \tool's multi-language support, users only need to use different frontends offered by srcML to create XML files or to extend srcML with new language parsers.

\vspace{-.5em}
\subsection{Program Encoding}\label{sec:phase2}
In an AST, the \textbf{depth} of root node is 0, all children of the root have depth=1, and grandchildren have depth=2. To create a program encoding (i.e., enhanced token sequence) for each candidate reference $p_j\in P$ or $p_i$, \tool first orders AST nodes based on depth-first traversal, counts the depth difference between any two adjacent leaf nodes (i.e., tokens), and injects related numbers of arrows (i.e., $\uparrow$ or $\downarrow$) into the original token sequence to reflect depth changes. 
As shown in Fig.~\ref{fig:lcs-optimize}, circled numbers \textcircled{1}-\textcircled{11} indicate traversal ordering. Between Nodes   \textcircled{3} and \textcircled{5}, because the depth is incremented from 2 to 3, a down arrow $\downarrow$ is injected between tokens \codefont{\#} and \codefont{include}. Notice that \textcircled{6} and \textcircled{8} have the same depth 2. As they belong to distinct \codefont{\#include} directives (i.e., program statements), \tool injects an extra pair of arrows ``$\uparrow\downarrow$'' between the tokens \codefont{<cstdio>} and \codefont{\#}, to indicate statement-level boundaries and penalize any potential token matches spanning across statements. 

\vspace{-.5em}
\subsection{Code Normalization}\label{sec:phase3}
Given programs $p_i$ and $p_j\in P$, there can be multiple LCSs with the same length.
Selecting the best one among all alternatives is crucially important, because it decides how well \tool matches code and derives differences. For our example in Section~\ref{sec:example}, there are  three distinct ways to match \codefont{\#include}-directives between $p_i$ and $p_j$. As shown in Fig.~\ref{fig:alts}, the three tokens of $p_i$ can either match the second directive of $p_j$ or match parts of $p_j$'s both directives. Within these three alternatives, the first one is most reasonable because the matching part does not span across statements. 
In this section, we first introduce our novel matching algorithm that identifies the optimal LCS to minimize statement-level differences.  
Next, we describe the program transformation that leverages optimal LCS to normalize each candidate reference $p_j$ to $p_j'$.

\subsubsection{The Design of Matching Algorithm} 
Our algorithm starts with the LCS mentioned in Section~\ref{sec:phase1}, 
enumerates all same-length LCSs, relies on the program encodings mentioned in Section~\ref{sec:phase2} to compute \textbf{cost} for each alternative, and picks the one with minimal cost. 
Here, the cost of an LCS alternative counts the number of arrows (i.e., depth changes) included by all consecutive fragments of \emph{unmatched} tokens. 
For instance, as shown in Fig.~\ref{fig:lcs-optimize}, suppose that we want to find the best LCS in $s$ to match a given three-token list $tl$: ``\codefont{\# include <iostream>}''. Sequence $s$ corresponds to two AST subtrees, with each token being a leaf node; $s$ has three alternatives to match $tl$: alt1, alt2, and alt3.
In alt1, the first three tokens compose an unmatched consecutive fragment, and there are two arrows inside this fragment (as highlighted by \hl{yellow}). Thus, the cost is 2. Similarly, in alt2 (or alt3), because the unmatched fragment covers three arrows, the cost is 3. As alt1 has the lowest cost, it is chosen as the optima. 

Intuitively, the cost function is applied to program encodings. It measures distances between unmatched tokens on ASTs and reflects potential repair effort. Namely, the farther away unmatched tokens are from each other, the more likely that they scatter on different subtrees, the more statements may need to be revised for program repair, and the higher cost there can be. Because finding the optimal LCS with minimum cost is a combinatorial optimization problem, we invented a DP algorithm to realize the algorithm design. 

\subsubsection{The Implementation of Matching Algorithm}
Our algorithm has three steps (see Algorithm~\ref{alg1}). 
Given a token sequence under comparison (i.e., $s_i$ or $s_j$) and the na\"ively extracted LCS $a\_lcs$ by Phase I (see Section~\ref{sec:phase1}), \tool first initializes two tables: $cost$ and $prev$. The $cost$ table records the minimum matching cost between any two subsequences; the $prev$ table traces the selected tokens in $s$ that lead to optimal solutions to subproblems. Next, \tool enumerates each token pair between $s$ and $a\_lcs$ to iteratively compute the minimum cost of each subproblem, and to update $prev$ accordingly. 

When comparing $s_i$ and $s_j$ for the optimal LCS, \tool applies Algorithm~\ref{alg1} twice: once to $(a\_lcs, s_i)$ and once to $(a\_lcs, s_j)$. It records the best match as a set of token pairs $M=\{m_1, m_2, \ldots, m_n\}$, where each pair $m_i (i\in[1, n])$ has a token from $s_i$ and a token from $s_j$. 
The time complexity of our algorithm is at most $O(len(s_i)*len(s_j))$. This new algorithm is much faster than the fastest tree matching algorithm (by Zhang and Shasha~\cite{Zhang-shasha89}), whose complexity is $O(len(s_i)^2 * len(s_j)^2)$. Such complexity contrast motivated us to compare programs based on token sequences instead of AST structures.


\input{alg1}

\subsubsection{Program Transformation}
\tool transforms each selected correct program $p_j\in P$ in order to (1) normalize the usage of variable/function names and (2) prepare candidate references for repair generation. 
Specifically, among the above-mentioned matching pairs $M$, \tool extracts all identifier mappings $IM (IM\subset M)$ and counts the occurrences of each unique match. 
When an identifier $id$ from one program is mapped to multiple identifiers $\{mid_1, mid_2, \ldots, mid_r\}$ in the other program, \tool computes the confidence value for each alternative pair as below: 

\small
\begin{equation*} 
Conf(mid_u) = \frac{count(mid_u)}{\sum_{v=1}^{r}count(mid_v)} \text{ where }u\in[1, r]
\end{equation*}
\normalsize 
If an alternative $mid_u$ has the confidence value no smaller than 0.6 and occurs at least 3 times, \tool keeps $mid_u$ and removes mappings conflicting with it.  Otherwise, \tool removes all alternatives from $IM$ and does not consider $id$ to be matched. Finally, based on the refined mappings $IM$, \tool converts $p_j$ to $p_j'$ by consistently replacing mapped identifiers. 

We chose the thresholds (0.6 and 3) based on two intuitions. First, if $p_i$ and $p_j$ are syntactically similar, most mappings in $IM$ should be consistent with each other; we can rely on the majority of mappings to align code and transform $p_j$. Thus, 0.6 is used to quantify the majority.   
Second, if $p_i$ and $p_j$ are syntactically different, their identifier mappings can conflict with each other and each mapping may occur only once or twice. To avoid any mismatches caused by such infrequent and unreliable token mappings, we used 3 to limit the minimum occurrence count for each selected token mapping. 
Our thresholds can impact \tool's applicability and repair quality. Although we did not explore other values, our evaluation (see Section~\ref{sec:evaluation}) shows that the used values worked well and enabled \tool to often suggest best repairs.

\vspace{-.5em}
\subsection{Repair Generation}
\label{sec:phase4}
This section first explains the extraction of candidate repairing edits, and then discusses repair minimization. 

\subsubsection{Extracting Candidate Repairing Edits}\label{sec:phase41}
\tool executes each normalized program with tests. If any normalized program $p_j'$ passes all tests, \tool uses $p_j'$ to generate candidate repairing edits $R_j$. Otherwise, if $p_j'$ fails any test while $p_j$ does not, it is possible that the identifier mappings \tool used to transform code are not fully accurate; \tool discards such normalized programs. 
When there are multiple sets of candidate edits available, \tool picks the one with fewest edits and we denote it as $R_s$.

Specifically for any code $p_j'$ that passes tests, \tool extracts $p_j$'s code differences with $p_i$ based on the optimal LCS mentioned in Phase III. 
Intuitively, the extracted differences reflect all edits required to transform $p_i$ into $p_j'$. 
Each edit may insert, delete, or update a sequence of tokens at a code location. 
One technical challenge here is: \emph{when \tool extracts LCS between program token sequences, it sometimes produces trivial mappings between the code snippets that should not have been matched.} Look at the following two snippets:

$p_i$: \codefont{\underline{x} == 0 || \underline{y} == \underline{0}}

$p_j'$: \codefont{a [ \underline{x} ] [ \underline{y} ] > \underline{0}} 

\noindent
Although the snippets are dissimilar to each other, \tool still manages to extract the LCS as ``\codefont{x y 0}''. When na\"ively relying on these trivial token mappings to generate edits, \tool can create
three edits that are less intuitive to users: (i) insert ``\codefont{a [}'' at location 0; (ii) update ``\codefont{== 0 ||}'' to ``\codefont{] [}'' at location 2; (iii) update ``\codefont{==}'' to ``\codefont{] >}'' at location 6. 
To improve the usefulness and readability of suggested repairing edits, \tool replaces the small edits co-applied to the same AST subtree with one big edit.
With more details, \tool first maps all edit locations to the corresponding leaf nodes on $p_i$'s AST. It then traverses the AST in a post-order way. 
If any subtree $st$ 
has more than 50\% of leaf nodes edited, then \tool replaces all related edits with one edit that updates the whole token sequence of $st$. For the example mentioned earlier in this section, the new edit used to replace all trivial edits (i.e., small and unintuitive edits) is
\begin{itemize}
\item[] Update ``\codefont{x == 0 || y == 0}'' to ``\codefont{a [ x ] [ y ] > 0}''
\end{itemize}
With such AST-based refinement, \tool can reduce the number of edits in some candidate repairs and remove meaningless edits. 
For our running example, \tool infers five edits: 

\begin{enumerate}
\item[1.] Insert ``\codefont{\# include <cstdio>}'' at location 0.
\item[2.] Delete ``\codefont{long}'' at location 13.
\item[3.] Delete ``\codefont{long}'' at location 21.
\item[4.] Update ``\codefont{a}'' to ``\codefont{a - 1}'' at location 56.
\item[5.] Update ``\codefont{>}'' to ``\codefont{>=}'' at location 59.
\end{enumerate}


\subsubsection{Repair Minimization}\label{sec:phase42} 
\tool computes the minimal repair by removing unneeded edits from $R_s$. 
The major challenge here is \emph{how to efficiently search for the minimal repair based on testing}. Although a na\"ive enumeration of all subsets of edits guarantees a minimum repair, it is almost inapplicable in practice. This is because given a repair with $n$ edits, the total number of subsets to explore is $(2^n - 2)$. This exponential search space 
 is intolerable, especially when $n$ is large (i.e., $n>10$) and every round of test-based validation is costly. 
To overcome the challenge, we invented a heuristic-based algorithm. Given a repair $R_s$ containing $n$ edits, our algorithm explores at most $(3n-1)$ subsets  in the following manner:

(1) \textbf{Single-edit repairs}: \tool applies each edit independently to $p_i$ and gets $n$ distinct versions. 
If any version passes all tests, the corresponding edit is the minimum repair.

(2) \textbf{Leave-one-out (LOO) repairs}: In each trial, 
\tool removes one edit from $R_s$ and applies the remaining $(n-1)$ edits to $p_i$. If any version passes all tests, the removed edit is an unneeded change and should be skipped. After these $n$ trials, \tool removes all unnecessary changes from $R_s$.

(3) \textbf{Leave-two-out (LTO) repairs}: Similar to LOO, \tool removes two consecutive edits in each trial and applies the remaining edits in $R_s$ to $p_i$. In this way, \tool filters out certain changes whose co-application does not affect the validation outcome. 

\begin{table}[h]
    \centering
    \scriptsize
    \caption{Minimizing the repair mentioned in Section~\ref{sec:phase41}}
     \label{tab:finalRepair}    
         \vspace{-.6em}
    \begin{tabular}{r| c|c}
        \toprule
        \textbf{Index} &\textbf{Edit Subset} & \textbf{Testing Outcome} \\\Xhline{2\arrayrulewidth}
        \hline
        1 & \{1\} & \textcolor{red}{\xmark} \\
        \hline
        2 & \{2\} & \textcolor{red}{\xmark} \\
        \hline
        3 & \{3\} & \textcolor{red}{\xmark} \\
        \hline
        4 &\{4\} & \textcolor{red}{\xmark} \\
        \hline
        5 &\{5\} & \textcolor{red}{\xmark} \\
        \hline
        6 &\{2, 3, 4, 5\} &\textcolor{blue}{\cmark} \\
        \hline
        7 & \{1, 3, 4, 5\} & \textcolor{blue}{\cmark} \\
        \hline
        8 & \{1, 2, 4, 5\} & \textcolor{blue}{\cmark}\\
        \hline
        9 & \{1, 2, 3, 5\} & \textcolor{red}{\xmark}\\
        \hline
        10 & \{1, 2, 3, 4\} & \textcolor{red}{\xmark}\\        
        \bottomrule
    \end{tabular}
        \vspace{-1.5em}
\end{table}
\noindent
Given the five edits mentioned in Section~\ref{sec:phase41} (see Table~\ref{tab:finalRepair}), \tool first applies each edit individually to $p_i$. Because no modified version passes all tests, there is no single-edit fix to $p_i$. Next, \tool applies four edits at a time. As edits $\{2, 3, 4, 5\}$ produce a correct program, the $1^{st}$ edit is unneeded. Similarly, we can remove the $2^{nd}$ and $3^{rd}$ edits.  
After LOO, $R_s=\{4,5\}$. There is no need to try LTO because only two edits are left. 

This algorithm was designed based on our manual analysis of randomly sampled solutions. We observed that (1) many incorrect solutions are similar to correct ones but only different in one or two places, and (2) each suggested edit is usually independent of others or at most related to another one edit. 
By limiting subset exploration to a small number of trials, our algorithm ensures \tool to always suggest repairs in a timely manner. In the worst case, even if an incorrect program does not match our observation, \tool can still suggest correct repairs although the repair sizes may be not optimal. 
Our later evaluation (see Section~\ref{sec:evaluation}) shows that \tool repaired many programs and had low runtime overhead; the suggested repairs were usually considered minimal by users. 

%% file: alg1.tex

\begin{algorithm}
\caption{Our novel DP algorithm to optimize LCS extraction}
\label{alg1}

\scriptsize
\KwIn{$s$, $a\_lcs$ \tcc{$s$ is $s_i$ or $s_j$, and $a\_lcs$ is the na\"ively extracted LCS by Phase I (see Section~\ref{sec:phase1})}}
\KwOut{$opt\_lcs$ \tcc{The optimal LCS from $s$}}
\tcc{1. Create two tables to separately save the matching cost and trace matched tokens}
$int$ $cost[len(a\_lcs)][len(s)]$; \\
$int$ $prev[len(a\_lcs)][len(s)]$;\\
\tcc{2. Enumerate each token pair between $s$ and $a\_lcs$ to compute the cost iteratively}
\ForEach{$q \in [0, len(s)-1]$}{
  \If{$match(a\_lcs[0], s[q])$}{
    $cost[0][q] = getCost(0, q-1)$;
  }\Else{ 
    $cost[0][q] = +\infty$;
  }
}
\ForEach{$p \in [1, len(a\_lcs)-1]$}{
  \ForEach{$q \in [1, len(s)-1]$}{
    \If{$match(a\_lcs[p], s[q])$}{
      \tcc{We update min while looping $q$, to reduce the overall time complexity to $O(len(s)*len(a\_lcs))$}
      $cost[p][q] = min(cost[p-1][k] + getCost(k+1, q-1))$; \tcc{$k$ varies within $[0, q-1)$}
      $prev[p][q]=k$ \tcc{record the $k$-value leading to the minimum cost}
    }\Else{
      $cost[p][q]=+\infty$;
    }
  }
}
$cost[len(a\_lcs)][len(s)]=min(cost[len(a\_lcs)-1][k]+getCost(k+1, len(s)-1))$ \tcc{$k\in[0, len(s)-1]$}
$prev[len(a\_lcs)][len(s)]=k$ \tcc{record the $k$-value related to the minimum cost}
\tcc{3. Traverse $prev$ to reveal the optimal LCS}
$opt\_lcs=getTokenLoc(prev)$;
\end{algorithm}

%% file: implement.tex
\section{Implementation Specifics}
\tool uses srcML~\cite{srcml}---a lightweight, highly scalable, robust, multi-language parsing tool to parse code and represent ASTs with XML files. \tool compares code based on the XML format, modifies code based on token sequences, and validates code modification via testing; thus, it has very weak dependence on language-specific features or program analysis frameworks. As srcML supports multiple programming languages, we implemented \tool to process both C and C++ program by using different srcML frontends. Based on our experience, \tool's implementation does not change with the language it processes, which indicates the tool's good portability across languages. Additionally, 
we used multithreading to speed up code comparison for Phase I.

%% file: evaluation.tex
\section{Evaluation}\label{sec:evaluation} 
We conducted two experiments with \tool and Clara to explore four research questions:
\begin{itemize}
\item \textbf{RQ1}: How effectively does \tool suggest repairs for incorrect solutions?
\item \textbf{RQ2}: How efficiently does \tool generate repairs?
\item \textbf{RQ3}: What is the quality of repairs generated by \tool?
\item \textbf{RQ4}: How does \tool compare with Clara in terms of the factors mentioned in RQ1-RQ3?
\end{itemize}
Our first experiment quantitatively measures the effectiveness and efficiency of \tool and Clara to answer RQ1, RQ2, and part of RQ4. The second experiment qualitatively measures the relevance, usefulness, size, and readability of repairs output by both tools to answer RQ3 and part of RQ4. All experiments were performed on a server with 32 Xeon E5-2667 3.2GHz processors. We set the process pool size to 8. 
This section introduces our datasets (Section~\ref{sec:datasets}) and evaluation metrics (Section~\ref{sec:metrics}), presents the results by \tool (Section~\ref{sec:toolresult}) and by Clara (Section~\ref{sec:clararesult}), and explains the comparison between \tool and Clara (Section~\ref{sec:comparison}). 

\vspace{-.5em}
\subsection{Datasets}
\label{sec:datasets}
We have two datasets: one dataset (\textbf{D1}) used to evaluate the effectiveness and efficiency of repair generation, and the other one (\textbf{D2}) used to evaluate the quality of generated repairs. 

\subsubsection{Dataset D1}
\tool handles C and C++ programs, so we collected the C/C++ program data from an online judge system---OpenJudge~\cite{openjudge}. 
OpenJudge is a platform where administrators post programming problems and users submit their solutions for evaluation. OpenJudge compiles each code submission, and runs the compiled version with predefined test cases. 
OpenJudge outputs ``\textbf{accepted (AC)}'' for any program passing all tests, or ``\textbf{wrong answer (WA)}'' for incorrect code that fails any test. 
Every user can submit multiple solutions for a problem. For each code submission, OpenJudge records the program, its judgment (WA or AC), the related problem Id, and the author Id. 
We chose five problems (see Appendix A in the supplementary document) from OpenJudge, because they were used in the course ``Introduction to Computing''. We downloaded all code submissions from OpenJudge. Since the program corpus is too large to use, we took two strategies when building D1 based on the corpus.

\begin{table}
\centering
\scriptsize
\caption{The C and C++ programs included in D1}
    \label{tab:d1}
\vspace{-.6em}
\begin{tabular}{C{.9cm}|C{.4cm}|R{.6cm}|R{.8cm}|R{1.cm}|R{1.cm}|R{1.cm}}
\toprule
 \textbf{Problem Id} & \textbf{\# of Tests} & \textbf{Language} & \textbf{LOC Range} &\textbf{Median LOC} & \textbf{\# of Correct} & \textbf{\# of Incorrect} \\ \Xhline{2\arrayrulewidth}
\multirow{2}{*}{1678} & \multirow{2}{*}{10}  
        & C &1 -- 65 & 18& 3,000 & 847\\ \cline{3-7}
     & & C++ &3 -- 416 &18 & 3,000 & 2,000 \\ 
\Xhline{2\arrayrulewidth}
\multirow{2}{*}{1689} & \multirow{2}{*}{10}
        & C & 1 -- 119& 15& 3,000 & 1,236 \\ \cline{3-7}
     & &C++ & 2 -- 448& 15&3,000 & 2,000 \\ 
\Xhline{2\arrayrulewidth}
\multirow{2}{*}{1703} & \multirow{2}{*}{10}
        & C &1 -- 609 & 17 &2,221 & 1431 \\\cline{3-7}
      &&C++ &2 -- 217 & 18& 3,000 & 2,000 \\        
\Xhline{2\arrayrulewidth}
\multirow{2}{*}{1716} & \multirow{2}{*}{10}
       & C &1 -- 177 &14&3,000 & 1,908  \\ \cline{3-7}
     && C++ &3 -- 164 &15& 3,000 & 2,000 \\ 
\Xhline{2\arrayrulewidth}
\multirow{2}{*}{1720} &\multirow{2}{*}{5}
       & C & 1 -- 57 &11& 3,000 & 1,254  \\\cline{3-7}
     && C++ & 2 -- 414 &11&3,000 & 2,000 \\
\bottomrule       
\end{tabular}     
\vspace{-1.5em}   
\end{table}

\textbf{\emph{Strategy 1}}: As a user may submit incorrect solutions for a problem before submitting a correct one, those incorrect solutions can be very similar to the correct one written by the same person. We filtered out a user's correct solution(s) if the user also submitted incorrect code. In this way, we ensure that (1) there is no data bias towards the correct/incorrect solutions written by the same person and (2) our experiment simulates \tool's real application scenario.

\textbf{\emph{Strategy 2}}: For every problem and each language, we randomly selected 3,000 correct programs and 2,000 incorrect programs to include in D1. When OpenJudge has insufficient program data for a particular problem or language, we simply included all available data after applying Strategy 1.

As shown in Table~\ref{tab:d1}, the adopted languages and problem Ids divide the whole dataset into 10 distinct program sets. The code size varies from 1 to 609 lines of code (LOC). 
Each program set includes 2,221-3,000 correct solutions and 847-2,000 incorrect ones. There are 5--10 test cases predefined for each problem.

\subsubsection{Dataset D2}
We created D2 based on the outputs by \tool and Clara. Specifically, among the incorrect solutions repaired by both tools, we randomly sampled 50 solutions in each C program set because Clara repairs C and Python code. Then we included the 250 sampled code (i.e., 50 * 5) together with tool-generated repairs into D2. 

\vspace{-.5em}
\subsection{Metrics}
\label{sec:metrics}
For RQ1, we defined two metrics: coverage and accuracy.

\emph{\textbf{Coverage (C)}} measures among all given incorrect programs, for how many of them a tool can suggest repairs: 

\small
\begin{equation*}
C = \frac{\text{\# of solutions with repairs generated}}{\text{Total \# of incorrect solutions}}\times 100\% 
\end{equation*}
\normalsize 
Coverage varies within [0\%, 100\%]. It reflects a tool's applicability. Namely, the higher coverage a tool achieves, the more diverse programs this tool can handle.

\emph{\textbf{Accuracy (A)}} measures among all suggested repairs, how many repairs lead to correct code that passes all tests. 

\small
\begin{equation*}
A = \frac{\text{\# of repaired programs that pass all tests}}{\text{\# of suggested repairs}}\times 100\% 
\end{equation*}
\normalsize 
Accuracy varies within [0\%, 100\%]. It measures the quality of all suggested repairs purely based on testing. Intuitively, if all repairs suggested by a tool pass all tests, the measured accuracy is 100\%. 

For RQ2, we measured tools' runtime overheads. For each tool, there are two kinds of cost. The first cost is \textbf{preprocessing time}, which measures how much time a tool spends on initialization and preparation. It corresponds to the time spent by \tool to create XML files based on AST parsing, and that spent by Clara to cluster correct solutions based on dynamic program analysis. The second cost is \textbf{average repair time}, which measures the average amount of time a tool spends on each repair suggestion. It is the time spent by \tool to derive and test repairs, and that by Clara to generate any repair. 
For RQ3, we defined two metrics to measure repair quality based on humans' manual inspection: relevance and usefulness. 

\emph{\textbf{Repair Relevance (RR)}} measures how relevant a repair is to the expected repair in instructors' mind. An instructor or TA grades a repair on a 1-4 scale as below:
\begin{itemize}
\item[\textbf{4:}] The repair is correct and fixes the bug(s), without any extra change.
\item[\textbf{3:}] The repair is correct and fixes the bug(s), but includes certain unnecessary change(s).
\item[\textbf{2:}] The repair is correct but loosely related to the original algorithm design. Namely, it applies so many edits that the original code is radically changed. 
\item[\textbf{1:}] The repair is incorrect.
\end{itemize}
$RR$ reflects the correctness of a suggested repair. The higher $RR$ is, the better. $RR$ is closely related to the accuracy metric $A$ mentioned above;  however, it measures the repair quality based on teachers' perception instead of testing results. 

\emph{\textbf{Repair Usefulness (RU)}} measures the usefulness of a repair suggestion perceived by teachers when they are given the repair together with the related incorrect solution. An instructor or TA grades a repair with a five-level Likert scale: 
\begin{itemize}
\item[\textbf{5:}] The repair suggestion is very useful.
\item[\textbf{4:}] The repair suggestion is somewhat useful.
\item[\textbf{3:}] The repair suggestion is neither useful nor useless.
\item[\textbf{2:}] The repair suggestion is less useful.
\item[\textbf{1:}] The repair suggestion is least useful.
\end{itemize}
$RU$ varies within [1, 5]. The higher score, the better.

For RQ4, in addition to the above-mentioned metrics, we designed and used two survey questions to query people's opinions on the repair comparison between \tool and Clara. 

\emph{\textbf{Size Comparison (SC).}} 
Given an incorrect solution and two repairs separately generated by \tool and Clara, participants are asked to choose among the following three options:

\begin{itemize}
\item[\textbf{A.}] The correct repairs suggested by \tool are smaller.
\item[\textbf{B.}] The correct repairs suggested by Clara are smaller.
\item[\textbf{C.}] The two types of repairs have almost equal sizes.
\end{itemize}

\emph{\textbf{Readability Comparison (RC).}} 
Given an incorrect solution and repairs suggested by tools, participants are asked to choose among the following three options:

\begin{itemize}
\item[\textbf{A.}] \tool's repair is easier to accept and understand.
\item[\textbf{B.}] Clara's repair is easier to accept and understand.
\item[\textbf{C.}] The two repairs are almost equally easy/difficult to accept and understand.
\end{itemize}
In the user study, participants did not know which tool is ours.
\vspace{-.5em}
\input{evaluation-fapr}

\vspace{-.5em}
\input{evaluation-clara}

\vspace{-.5em}
\input{evaluation-comparison}

%% file: evaluation-fapr.tex
\subsection{Experiments with \tool}
\label{sec:toolresult}
In the first experiment, we applied \tool to D1. In the second experiment, we recruited 
five CS students to separately inspect the sampled programs and related \tool's repairs in D2, assuming them to play the TA role.
 These students had 2-8 years of C experience; each of them
 independently assessed the repair quality for samples from one C program set. 

 
\begin{table}
\centering
\scriptsize
\caption{\tool's coverage, accuracy, and runtime overhead}\label{tab:rq1-fapr}
\vspace{-.6em}
\begin{tabular}{C{1.2cm}|C{1.2cm}|R{.6cm}|R{.6cm}|R{1cm}|R{1.4cm}}
    \toprule
    \textbf{Language} & \textbf{Problem} & \textbf{C} & \textbf{A} & \multicolumn{2}{c}{\textbf{Runtime Overhead (second)}} \\ \cline{5-6}
    & \textbf{Id} & \textbf{(\%)} & \textbf{(\%)} & \textbf{Preprocessing}& \textbf{Average Repair Time}\\\Xhline{2\arrayrulewidth}
\multirow{5}{*}{C} & 1678 & 95.9& 100& 65.8 &4.1 \\ \cline{2-5}
			    & 1689 & 99.9& 100& 65.5 & 3.8\\ \cline{2-5} 
			    & 1703 & 99.5& 100& 48.7 & 2.9\\ \cline{2-5}
			    & 1716 & 99.1& 100& 64.4 & 2.6\\ \cline{2-5}
			    & 1720 & 99.4& 100& 61.9 &1.4\\ \Xhline{2\arrayrulewidth}
\multirow{5}{*}{C++} & 1678 & 95.5& 100& 66.8 & 8.1 \\ \cline{2-5}
			       & 1689 & 99.4& 100& 65.6 &5.4\\ \cline{2-5} 
			       & 1703 & 99.8& 100&  65.5 &5.2\\ \cline{2-5}
			       & 1716 & 99.2& 100& 63.2 &4.4\\ \cline{2-5}
			       & 1720 & 99.4& 100& 62.6 &3.6\\\bottomrule		    
\end{tabular}
\vspace{-1.5em}
\end{table}

\subsubsection{Coverage} 
As shown in Table~\ref{tab:rq1-fapr}, \tool achieved very high coverage. It suggested repairs for 95.9\%--99.9\% of incorrect C programs; it obtained 95.5\%--99.8\% coverage when being applied to 5 C++ program sets. 
We were curious why \tool was unable to suggest repairs for some programs, so we manually inspected code and found two reasons. 
First, some programs contain special characters (e.g., Chinese characters) that prevents srcML from parsing code. Second, some incorrect solutions are dissimilar to all correct ones and the normalized programs all failed test(s), being unusable to repair code. In the future, we will overcome the first limitation by preprocessing code to remove special characters, and mitigate the second one by enlarging the corpus of correct code.


\subsubsection{Accuracy}
In all program sets, \tool achieved 100\% accuracy. It means that the suggested repairs by \tool always pass all tests. This is expected, because we designed \tool to validate each generated repair via testing. If \tool cannot produce any repair to fix a given program, the program is not considered in the accuracy evaluation. 

\subsubsection{Runtime Overhead} As shown in Table~\ref{tab:rq1-fapr}, in each program set, the preprocessing time of converting all correct code to XML files is 48.7-66.8 seconds while the average repair time is 1.4-8.1 seconds. Our evaluation implies that when teachers rely on \tool to locate bugs and suggest repairs, \tool can always respond quickly.

\vspace{0.2em}
\noindent\begin{tabular}{|p{8.4cm}|}
	\hline
	\textbf{Finding 1:} \emph{Our quantitative experiment with \tool shows that it generated repairs with very high coverage ($\ge 95.5\%$), 100\% accuracy, and low runtime overheads ($\le 8.1$ seconds per repair)}.
	\\
	\hline
\end{tabular}
\vspace{0.3em}

\subsubsection{Repair Relevance} 
As shown in the left part of Table~\ref{tab:rr}, among the 250 samples, participants considered 224 repairs to be correct and include no unnecessary change.
They identified another 15 repairs to be correct but include unnecessary edits. Another seven repairs were graded to be ``2''; they were considered to be correct but loosely related to the original program. Four repairs were graded to be ``1'' and considered incorrect, which is unexpected to us because \tool only suggested repairs after programs passed all tests. We inspected the reported incorrect repairs, and found the predefined test cases to have limited branch coverage. Consequently, some adopted ``correct'' programs pass all test cases but actually contain bugs, which make generated repairs incorrect. 

\begin{table}
\centering
\caption{The repair relevance (RR) of \tool and Clara}
\label{tab:rr}
\scriptsize
\vspace{-.6em}
\begin{tabular}{C{1.5cm}?r|r|r|r? r|r|r|r}
\toprule
\textbf{Problem} & \multicolumn{4}{c?}{\textbf{\tool}} & \multicolumn{4}{c}{\textbf{Clara}} \\ \cline{2-9}
\textbf{Id} & \textbf{4} & \textbf{3} & \textbf{2} & \textbf{1} & \textbf{4} & \textbf{3} &\textbf{2} & \textbf{1}\\ \Xhline{2\arrayrulewidth}
1678 & 43 & 3 & 1 & 3 & 24  & 11 &0  &15 \\ \hline
1689 & 45 & 4 & 1 & 0 & 25 & 16 & 3 & 6 \\ \hline
1703 & 43 & 4 & 3 & 0 & 9 & 17 & 0 & 24\\ \hline
1716 & 50 & 0 & 0 & 0 & 11 & 8 & 0 & 31 \\ \hline
1720 & 43 & 4 & 2 & 1 & 27 & 12 &7 & 4 \\ 
 \Xhline{2\arrayrulewidth}
\textbf{Total} & 224 & 15 & 7 & 4 & 96& 64 & 10 &80 \\ \bottomrule
\end{tabular}
\end{table}

\begin{table}
\scriptsize
\vspace{-.6em}
\caption{The repair usefulness (RU) of \tool and Clara}
\label{tab:ru}
\vspace{-.6em}
\begin{tabular}{C{1.2cm}?r|r|r|r|r? r|r|r|r|r}
\toprule
\textbf{Problem} & \multicolumn{5}{c?}{\textbf{\tool}} & \multicolumn{5}{c}{\textbf{Clara}} \\ \cline{2-11}
\textbf{Id} & \textbf{5} & \textbf{4} & \textbf{3} & \textbf{2} & \textbf{1} &\textbf{5} &\textbf{4} & \textbf{3} &\textbf{2} & \textbf{1}\\ \Xhline{2\arrayrulewidth}
1678 & 47 & 2 & 0 & 0 & 1 &35 & 2&4&3&6\\ \hline
1689 & 49 & 1 & 0 & 0 & 0 &40&2&1&1&6\\ \hline
1703 & 44 & 5 & 1 & 0 & 0 &14&10&2&0&24\\ \hline
1716 & 50 & 0 & 0 & 0 & 0 &12&10&10&2&16\\ \hline
1720 & 43 & 6 & 1 & 0 & 0 &34&7&7&0&2 \\ \Xhline{2\arrayrulewidth}
\textbf{Total} & 233 & 14 & 2 &0 & 1&135&31&24&6&54 \\
\bottomrule
\end{tabular}
\vspace{-1.5em}
\end{table}

\subsubsection{Repair Usefulness}
According to the left part of Table~\ref{tab:ru}, participants believed 247 out of 250 repair samples to be very useful or somewhat useful. Only two repairs were graded to be ``neither useful nor useless'' and one repair was considered ``useless''. We further inspected the three less helpful repairs to understand why participants disliked them. The one considered useless is an incorrect repair \tool suggested due to limited test coverage; the other two significantly changed the algorithm design. 
\tool suggested such algorithm-modifying repairs, as the adopted correct code implements different algorithms from the original incorrect code. 
Participants prefer repairs that fix bugs instead of replacing whole algorithms. 

\vspace{0.5em}
\noindent\begin{tabular}{|p{8.4cm}|}
	\hline
	\textbf{Finding 2:} \emph{Our qualitative experiment with \tool's repairs shows that participants considered (1) 89.6\% of repairs to be correct and smallest and (2) 98.8\% of repairs to be helpful. The average RR score is 3.8 out of 4, and the average RU score is 4.9 out of 5.}
	\\
	\hline
\end{tabular}

%% file: evaluation-clara.tex
\subsection{Experiments with Clara}
\label{sec:clararesult}

In the first experiment, we applied Clara to 5 C program sets in D1 because Clara does not repair C++ code. In the second experiment, similar to what we did in Section~\ref{sec:toolresult}, we also asked the five participants to manually inspect Clara's repairs for the sampled incorrect solutions in D2. Namely, given an incorrect solution, a participant was expected to inspect the alternative repairs suggested by both tools without knowing which tool is ours. With such experiment setting, we ensured that the alternative repairs for the same incorrect solution were always examined by the same person and got assessed based on the same subjective criteria. 
Section~\ref{sec:toolresult} reports participants' assessment for \tool repairs, while this section reports participants' assessment for Clara repairs.

\begin{table}
    \centering
    \footnotesize
    \caption{Clara's coverage, accuracy, and runtime overhead}\label{tab:rq1-clara}
    \vspace{-.6em}
    \begin{tabular}{C{1.2cm}|R{1.2cm}|R{1.2cm}|R{1.cm}|R{1.8cm}}
    \toprule
    \textbf{Problem} & \textbf{Coverage} & \textbf{Accuracy} & \multicolumn{2}{c}{\textbf{Runtime Overhead (second)}}\\ \cline{4-5}
    \textbf{Id} & \textbf{(\%)} & \textbf{(\%))} & \textbf{Preprocessing}& \textbf{Average Time Per Repair}\\\Xhline{2\arrayrulewidth}
 1678 &83.2& -& 19,039&49.5 \\ \hline
 1689 & 57.1& -& 184,212&40.6\\ \hline
 1703 & 77.5& -& 5,094&29.4\\ \hline
 1716 & 83.2& -& 14,926&13.2\\ \hline
 1720 & 92.0& -& 7&40.8\\\bottomrule		    
 \multicolumn{5}{l}{``-'' mean that the data is unavailable.}
    \end{tabular}
    \vspace{-1.5em}
\end{table}

\subsubsection{The Clara Approach}
To automate repair generation, Clara first clusters correct solutions based on their semantic equivalence (e.g., equivalent variable states during program execution), and then compares each incorrect solution with the representative 1-2 solutions from each cluster to produce repairs. Specifically, to capture program semantics, Clara parses source code to create an internal representation (IR), and adopts an interpreter to execute each IR statement and record intermediate execution status (i.e., variable values). The recorded program states were later used to decide whether two programs implement the same semantics (i.e., by having equivalent variable values). Clara extracts IR differences for repairs. For example, given \codefont{for(int i=a;i>0;i$--$)} updated to \codefont{for(int i=a-1;i>=0;i$--$)}, Clara expresses the edit as: 

\begin{itemize}
\item[1.] Change \codefont{i := a} to \codefont{i := -(a, 1)} 
\item[2.] Change \codefont{\$cond := >(a, 0)} to \codefont{\$cond := >=(a, 0)} 
\end{itemize}

\subsubsection{Coverage} Table~\ref{tab:rq1-clara} shows the evaluation results for Clara's coverage, accuracy, and runtime overhead. According to the table, Clara achieved 57.1\%--92.0\% coverage when being applied to the 5 C program sets. To understand why Clara was unable to repair all programs, we manually analyzed its implementation. We found that Clara has a self-defined C parser for C-to-IR conversion, and a self-defined interpreter to execute IR instructions and produce traces for dynamic analysis. Both parts only handle general cases, but cannot process all syntactic structures or library function calls.  

\subsubsection{Accuracy} As mentioned in Clara's approach overview, the tool suggests repairs based on IR instead of C code; it does not transform IR back to C code to reflect the code modification, neither does it test code to automatically validate repairs. Thus, we could not measure the tool's accuracy via automated testing.

\subsubsection{Runtime Overhead} 
According to Table~\ref{tab:rq1-clara}, Clara's preprocessing time is 7--184,212 seconds. 
It spent only 7 seconds to cluster code for Problem 1720, mainly because the programs are very simple and produce a small number of intermediate states. Clara spent 184,212 seconds (51.2 hours) clustering code for Problem 1689, because the programs are more complicated and have many more intermediate states for Clara to trace and compare. Additionally, Clara spent 13.2--49.5 seconds on average to suggest each repair. 

\vspace{0.2em}
\noindent\begin{tabular}{|p{8.4cm}|}
	\hline
	\textbf{Finding 3:} \emph{Our quantitative experiment with Clara shows that it generated repair with good coverage (57.1\%--92.0\%), 
	but incurred relatively high runtime overheads for clustering 
	and repair suggestion.} 
	\\
	\hline
\end{tabular}
\vspace{0.3em}

\subsubsection{Repair Relevance} According to the right part of Table~\ref{tab:rr}, participants identified 96 repairs to be correct and minimized, and recognized another 64 repairs to be correct but have redundant changes. They also considered 10 repairs to be correct but weakly related to the original algorithm design, and 80 repairs to be incorrect. 

\subsubsection{Repair Usefulness} As shown by the right part of Table~\ref{tab:ru}, participants believed 166 out of 250 repairs to be useful and 60 repairs to be useless. They were neutral to the remaining 24 repairs. The participants did not like 84 (i.e., $60+24$) of the repairs because these repairs are either incorrect or totally replace the algorithm design. One possible reason to explain Clara's many unsatisfactory repairs is its semantics-based program comparison. By comparing and clustering programs based on equivalent execution status, Clara managed to align semantically similar but syntactically dissimilar code with each other. Although such flexible alignment can locate more candidate references to fix $p_i$, the generated repairs sometimes implement irrelevant algorithms and become less useful to participants.

\vspace{0.5em}
\noindent\begin{tabular}{|p{8.4cm}|}
	\hline
	\textbf{Finding 4:} \emph{Our qualitative experiment with Clara's repair samples shows that participants found (1) 38.4\% of repairs to be correct and smallest and (2) 66.4\% of repairs to be helpful. The average RR value is 2.7 out of 4, and the average RU value is 3.7 out of 5.}
	\\
	\hline
\end{tabular}

%% file: evaluation-comparison.tex
\subsection{Comparison between \tool and Clara}
\label{sec:comparison}
Observing the results shown in Sections~\ref{sec:toolresult} and~\ref{sec:clararesult}, we learnt that \tool outperformed Clara by achieving higher coverage (i.e., fixing more code) with lower runtime overheads, and producing repairs with higher RR and RU scores. 
Table~\ref{tab:compare} shows participants' manual comparison between the repairs of both tools. According to the table, when participants compared repair sizes, they considered 105 repairs by \tool to be smaller than Clara's repairs; however, only 12 of Clara's repairs were considered smaller. For 133 incorrect solutions, the repairs generated by both tools seem to have equal sizes. In terms of readability, participants considered 170 of \tool's repairs to be more readable; only 10 of Clara's repairs were considered to have better readability.

\begin{table}
\caption{Repair comparison between \tool and Clara}\label{tab:compare}
\vspace{-.6em}
\centering
\scriptsize
\begin{tabular}{C{1.2cm}? R{.6cm}|R{.6cm}|R{.6cm}? R{.6cm}|R{.6cm}|R{.6cm}}
\toprule
\textbf{Program} &\multicolumn{3}{c?}{\textbf{Size Comparison}} & \multicolumn{3}{c}{\textbf{Readability Comparison}}\\ \cline{2-7}
\textbf{Id} & \textbf{A} & \textbf{B} & \textbf{C} &\textbf{A} &\textbf{B}&\textbf{C} \\ 
\Xhline{2\arrayrulewidth}
1678 & 25 & 6 & 19 & 18 & 6 & 26 \\ \hline
1689 & 18 & 5 & 27 & 16 & 2 & 32 \\ \hline
1703 & 19 & 0 & 31 & 50 & 0 & 0 \\ \hline
1716 & 38 & 0 & 12 & 41 & 0 & 9 \\ \hline
1720 & 5 & 1 & 44 & 45 & 2 & 3 \\ \bottomrule
\textbf{Total} & 105 & 12 & 89 & 170 &10 &70\\ \bottomrule
\end{tabular}
\vspace{-1.5em}
\end{table}

Three reasons can explain \tool's higher efficiency and better repair quality. First, \tool compares code based on static token-level matches instead of semantic reasoning via dynamic analysis; its comparison method is simpler and thus faster. Second, participants preferred bug fixes over algorithm replacements; semantics-based code comparison does not ensure syntactic similarity and thus Clara's repairs are less accurate. Third, \tool suggests repairs in C, which enables automated validation via testing and facilitates participants to comprehend repairs. Additionally, \tool has better portability than Clara because it does not use complex program analysis. To apply \tool to code written in a new language, developers only need to feed \tool the XML files of program ASTs.

\vspace{0.5em}
\noindent\begin{tabular}{|p{8.4cm}|}
	\hline
	\textbf{Finding 5:} \emph{The comparison between tools shows that \tool worked better than Clara. \tool suggested correct repairs for more programs. In many scenarios, users considered \tool's repairs to have higher quality (i.e., smaller and more readable). }
	\\
	\hline
\end{tabular}

%% file: threats.tex
\section{Discussion}
To reveal the similarity or differences between programs, people can compare either code strings, token sequences, ASTs, control/data flows, program dependencies, or execution traces. Different comparison methods achieve distinct trade-offs among speed, flexibility, and cross-language portability. For instance, string comparison is usually fast and applicable to different languaged code, but does not tolerate any difference (e.g., formatting) when matching code. On the other extreme, execution comparison is usually slow and only applicable to languages with good dynamic analysis support; however, it can flexibly match programs that are syntactically different but semantically equivalent. In MOOC, lots of correct code is available and students' incorrect solutions in different languages wait for debugging feedback; we believe (1) speed and cross-language portability to be more important than flexibility and (2) syntax-driven approaches to outperform semantics-driven ones.

Instead of trying all edit subsets, \tool minimizes repairs by enumerating 
single-edit repairs, leave-one-out repairs, and leave-two-out repairs; it also relies on testing outcomes to skip the exploration of some unpromising subsets. We designed this algorithm based on our manual analysis of sample programs (see Section~\ref{sec:phase42}). In the future, we will explore more heuristics that help \tool reduce repairs more effectively  without introducing too much overhead.


%% file: related.tex
\section{Related Work}
The related work of our research includes 
automatic patch generation and automated feedback generation.

\paragraph*{Automatic Patch Generation}
Given a buggy program that fails some test(s) in a test suite, automatic program repair (APR) generates candidate patches and checks patch correctness via compilation and testing~\cite{Perkins09:clearview, Weimer2009:repair,Long2016:Prophet,Qi2014,DeMarco2014,Kim2013:PAR,LeGoues12:gp,Singh13,Nguyen13}. For example, GenProg~\cite{LeGoues12:gp} and RSRepair~\cite{Qi2014} create patches by randomly replicating, mutating, or deleting code from the existing program. As random search often has a huge search space, these tools have high runtime overheads but unsatisfactory capabilities. They 
can only fix a few kinds of bugs and the repairs are usually limited to single-line edits. 

Data-driven program repair tools extract edit patterns from frequently applied bug fixes, and use those patterns to fix similar buggy code~\cite{Meng11,Meng13,Bader19,Tufano19}. For instance, Getafix~\cite{Bader19} mines software repositories for code changes that fix a specific kind of bugs (e.g., NullPointerException) and generalizes edit patterns from those changes.
Given a previously unseen buggy program, Getafix finds suitable fix patterns based on the edit context, ranks all candidate fixes, and suggests the top fixes. Tufano et al.~\cite{Tufano19} applies Neural Machine Translation (NMT) to bug-fixing commits, 
to train a model that predicts repaired code given buggy programs. These tools are similar to \tool, but they do not guarantee that the suggested repairs pass all tests or always have correct syntax and semantics.

\begin{figure}
\centering
\includegraphics[width=.65\linewidth]{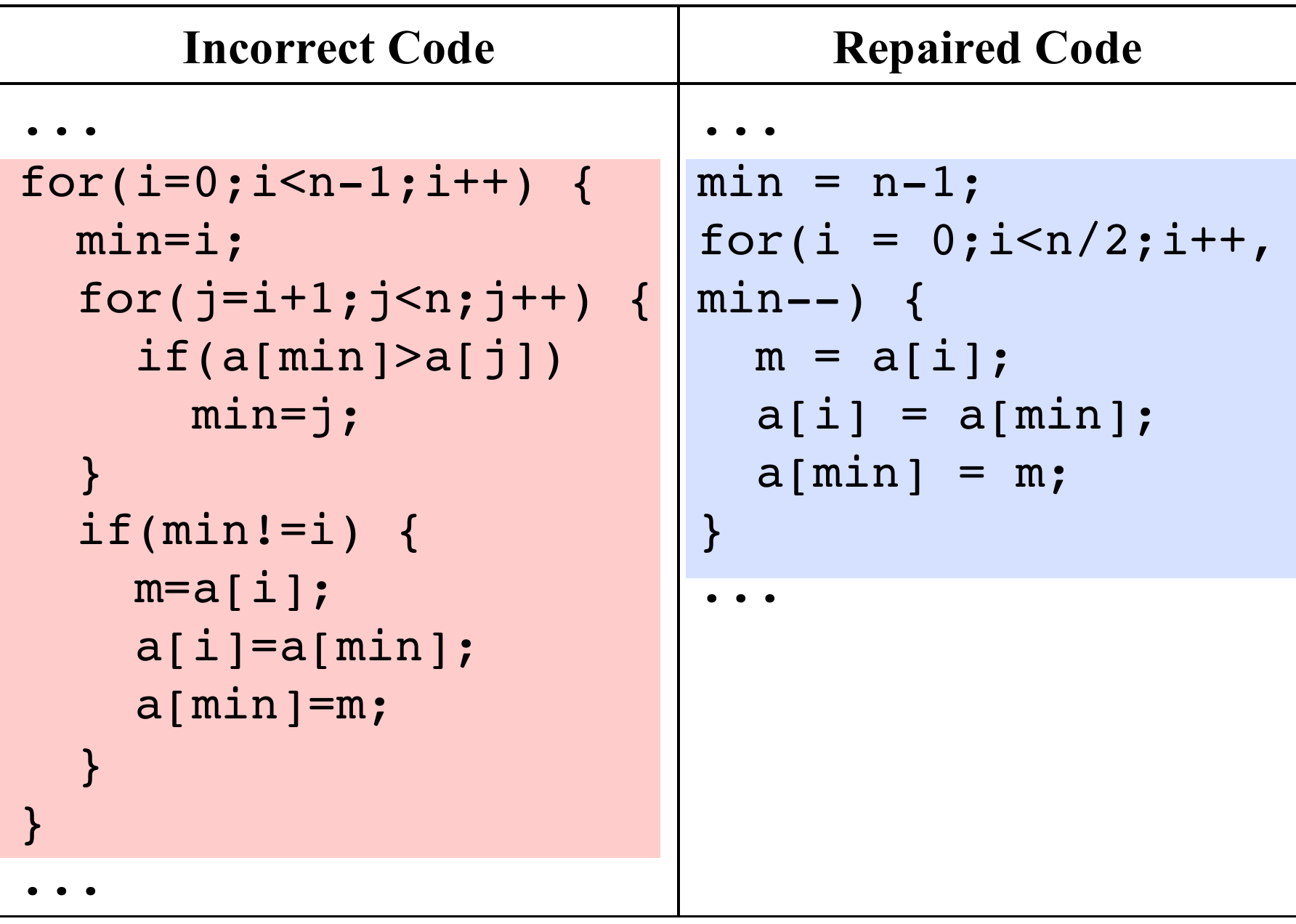}
\vspace{-.6em}
\caption{One of \tool's repairs that applies complex edits}\label{fig:ex2}
\vspace{-1.5em}
\end{figure}

Several tools use symbolic execution to reason about the semantics of both incorrect and correct programs, in order to repair code~\cite{Mechtaev16,Mechtaev18}. For instance, Angelix~\cite{Mechtaev16} first uses statistical fault localization to identify $n$ most suspicious expressions in a buggy program. It then relies on concolic execution to extract value constraints on expressions that should be satisfied by a repaired program, and adopts program synthesis to generate repairs accordingly. 
However, due to the scalability and applicability issues of symbolic execution, these approaches cannot create complex patches that edit large chunks of code. In comparison, \tool can create complex patches, as demonstrated by Fig.~\ref{fig:ex2}. Please refer to Appendix B in our supplementary document for more examples. 
\paragraph*{Knowledge-Based Feedback Generation}
Tools were proposed to generate feedback based on domain knowledge or predefined rules~\cite{oj,openjudge,leetcode,Isong01,Reek96,Jackson97,Luck99,Edwards03,Singh13}. 
Specifically, given program submissions, various online judge systems~\cite{oj,openjudge,leetcode} and automatic grading systems~\cite{Isong01,Reek96,Luck99} check for code correctness via compilation and testing. If a program compiles successfully and passes all predefined test cases, these systems consider the code to be correct; otherwise, they provide high-level succinct error description like ``Time Limit Exceeded (TLE)'' or ``Memory Limit Exceeded (MLE)''. Such vague and abstract program feedback may be not helpful to students. 

When using AutoGrader~\cite{Singh13} to repair students' incorrect code, instructors have to learn a domain-specific language EML (short for ``Error Model Language''), and use EML to describe possible coding errors and related fixes. Given a buggy program, AutoGrader matches the program with all error models, and enumerates related fixing patterns to correct the code. However, it is tedious and error-prone for instructors to learn EML and to prescribe error models, while the specified models can only fix a limited number of bugs. 
\tool is different from prior work in two aspects. First, it does not require instructors to prescribe error models and related fixing patterns. Second, it  generates repair based on existing correct solutions and predefined test cases. 

\paragraph*{Data-Based Feedback Generation}
Some tools were proposed to generate feedback based on the comparison between correct and incorrect solutions~\cite{Adam80,Tillmann13,Tillmann14,Zimmerman15,Kim16,Kaleeswaran16,pu2016skp,Wang18,Song19,Hu19,Birch:2019aa}. For instance, 
Pex4Fun \cite{Tillmann13} and CodeHunt~\cite{Tillmann14} use dynamic symbolic execution to generate test cases that reveal behavioral discrepancies between student solutions and the teacher's hidden program specification. Similar to Clara, Laura~\cite{Adam80} and Apex~\cite{Kim16} also model and compare program semantics to repair code. However, Laura statically analyzes code to represent semantics with graphs, which capture the value constraints and evaluation ordering between variables. Apex collects symbolic and concrete execution traces  to reason about program semantics, and uses Z3 to reveal semantic equivalence. Due to the usage of symbolic execution, dynamic analysis, and/or graph-based semantic reasoning, these tools cannot achieve high efficiency. \tool complements these tools.

Pu et al.~\cite{pu2016skp} built sk\_p---that trains generative neural networks with correct programs, and uses the trained model to suggest fixes for incorrect programs. Evaluation shows that sk\_p could not suggest fixes with high accuracy.
DrRepair~\cite{Yasunaga20} and DeepFix~\cite{Gupta17} also adopt machine learning methods to repair programs; however, they address compilation errors and cannot fix programs that fail tests. 
 SARFGEN~\cite{Wang18} is most relevant to our work. Given an incorrect program, it searches for syntactically similar correct code based on program embeddings, aligns code based on control-flow structures and tree edit distances, and generates repairs based on code differences. However, SARFGEN uses dynamic analysis to minimize repairs, which may jeopardize its cross-language portability and repair quality. As the tool is unavailable, we did not empirically compare \tool with SARFGEN.

%% file: conclusion.tex
\section{Conclusion}
Albeit similar to some prior work that compares incorrect programs against correct ones to create repairs, \tool is novel in three aspects. First, \tool retrieves correct code $P$ most similar to the given incorrect code $p_i$ based on token-level comparison. Such efficient comparison ensures the (1) syntactic similarity between $P$ and $p_i$, and (2) usability of $P$ to generate minimal repairs that do not considerably modify algorithm design. 
Second, \tool encodes syntactic structures into token sequences; such encodings help \tool overcome any inaccuracy caused by pure token-level comparison and improve the quality of generated repairs. Third, \tool minimizes repairs by selectively testing subsets of generated edits; this testing-based approach eliminates \tool's dependency on complex program analysis and ensures its cross-language portability. Thanks to the novelties mentioned above, \tool complements existing work and can suggest repairs in a fast and accurate way.    

Our comprehensive evaluation with 45,897 C/C++ programs evidences \tool's great efficiency, accuracy, and cross-language portability. \tool requires almost zero developer effort to get ported from C to C++ code. 
In the future, we will deploy \tool to online judge systems. 

%% file: yunlong-icsme-2021.bbl
\begin{thebibliography}{10}
\providecommand{\url}[1]{#1}
\csname url@samestyle\endcsname
\providecommand{\newblock}{\relax}
\providecommand{\bibinfo}[2]{#2}
\providecommand{\BIBentrySTDinterwordspacing}{\spaceskip=0pt\relax}
\providecommand{\BIBentryALTinterwordstretchfactor}{4}
\providecommand{\BIBentryALTinterwordspacing}{\spaceskip=\fontdimen2\font plus
\BIBentryALTinterwordstretchfactor\fontdimen3\font minus
  \fontdimen4\font\relax}
\providecommand{\BIBforeignlanguage}[2]{{%
\expandafter\ifx\csname l@#1\endcsname\relax
\typeout{** WARNING: IEEEtran.bst: No hyphenation pattern has been}%
\typeout{** loaded for the language `#1'. Using the pattern for}%
\typeout{** the default language instead.}%
\else
\language=\csname l@#1\endcsname
\fi
#2}}
\providecommand{\BIBdecl}{\relax}
\BIBdecl

\bibitem{mooc}
``{Remember the MOOCs? After Near-Death, They're Booming},''
  \url{https://www.nytimes.com/2020/05/26/technology/moocs-online-learning.html},
  2020.

\bibitem{Singh13}
\BIBentryALTinterwordspacing
R.~Singh, S.~Gulwani, and A.~Solar-Lezama, ``Automated feedback generation for
  introductory programming assignments,'' in \emph{Proceedings of the 34th ACM
  SIGPLAN Conference on Programming Language Design and Implementation}, ser.
  PLDI '13.\hskip 1em plus 0.5em minus 0.4em\relax New York, NY, USA:
  Association for Computing Machinery, 2013, pp. 15--26. [Online]. Available:
  \url{https://doi.org/10.1145/2491956.2462195}
\BIBentrySTDinterwordspacing

\bibitem{pu2016skp}
Y.~Pu, K.~Narasimhan, A.~Solar-Lezama, and R.~Barzilay, ``sk\_p: a neural
  program corrector for moocs,'' 2016.

\bibitem{Kaleeswaran16}
\BIBentryALTinterwordspacing
S.~Kaleeswaran, A.~Santhiar, A.~Kanade, and S.~Gulwani, ``Semi-supervised
  verified feedback generation,'' in \emph{Proceedings of the 2016 24th ACM
  SIGSOFT International Symposium on Foundations of Software Engineering}, ser.
  FSE 2016.\hskip 1em plus 0.5em minus 0.4em\relax New York, NY, USA:
  Association for Computing Machinery, 2016. [Online]. Available:
  \url{https://doi.org/10.1145/2950290.2950363}
\BIBentrySTDinterwordspacing

\bibitem{Gulwani18}
\BIBentryALTinterwordspacing
S.~Gulwani, I.~Radi\v{c}ek, and F.~Zuleger, ``Automated clustering and program
  repair for introductory programming assignments,'' in \emph{Proceedings of
  the 39th ACM SIGPLAN Conference on Programming Language Design and
  Implementation}, ser. PLDI 2018.\hskip 1em plus 0.5em minus 0.4em\relax New
  York, NY, USA: Association for Computing Machinery, 2018, pp. 465--480.
  [Online]. Available: \url{https://doi.org/10.1145/3192366.3192387}
\BIBentrySTDinterwordspacing

\bibitem{Perry19}
\BIBentryALTinterwordspacing
D.~M. Perry, D.~Kim, R.~Samanta, and X.~Zhang, ``Semcluster: Clustering of
  imperative programming assignments based on quantitative semantic features,''
  ser. PLDI 2019.\hskip 1em plus 0.5em minus 0.4em\relax New York, NY, USA:
  Association for Computing Machinery, 2019, pp. 860--873. [Online]. Available:
  \url{https://doi.org/10.1145/3314221.3314629}
\BIBentrySTDinterwordspacing

\bibitem{openjudge}
``{OpenJudge},'' \url{http://openjudge.cn}, 2020.

\bibitem{srcml}
``{srcML},'' \url{https://www.srcml.org}, 2020.

\bibitem{aho06}
A.~V. Aho, M.~S. Lam, R.~Sethi, and J.~D. Ullman, \emph{Compilers: Principles,
  Techniques, and Tools (2nd Edition)}.\hskip 1em plus 0.5em minus 0.4em\relax
  USA: Addison-Wesley Longman Publishing Co., Inc., 2006.

\bibitem{Zhang-shasha89}
\BIBentryALTinterwordspacing
K.~Zhang and D.~Shasha, ``Simple fast algorithms for the editing distance
  between trees and related problems,'' \emph{SIAM J. Comput.}, vol.~18, no.~6,
  pp. 1245--1262, Dec. 1989. [Online]. Available:
  \url{https://doi.org/10.1137/0218082}
\BIBentrySTDinterwordspacing

\bibitem{Perkins09:clearview}
J.~H. Perkins, S.~Kim, S.~Larsen, S.~Amarasinghe, J.~Bachrach, M.~Carbin,
  C.~Pacheco, F.~Sherwood, S.~Sidiroglou, G.~Sullivan, W.-F. Wong, Y.~Zibin,
  M.~D. Ernst, and M.~Rinard, ``Automatically patching errors in deployed
  software,'' in \emph{SOSP '09: Proceedings of the ACM SIGOPS 22nd symposium
  on Operating systems principles}.\hskip 1em plus 0.5em minus 0.4em\relax New
  York, NY, USA: ACM, 2009, pp. 87--102.

\bibitem{Weimer2009:repair}
W.~Weimer, T.~Nguyen, C.~Le~Goues, and S.~Forrest, ``Automatically finding
  patches using genetic programming,'' in \emph{ICSE '09: Proceedings of the
  31st International Conference on Software Engineering}.\hskip 1em plus 0.5em
  minus 0.4em\relax Washington, DC, USA: IEEE Computer Society, 2009, pp.
  364--374.

\bibitem{Long2016:Prophet}
F.~Long and M.~Rinard, ``Automatic patch generation by learning correct code,''
  \emph{SIGPLAN Not.}, 2016.

\bibitem{Qi2014}
Y.~Qi, X.~Mao, Y.~Lei, Z.~Dai, and C.~Wang, ``The strength of random search on
  automated program repair,'' in \emph{ICSE}, 2014.

\bibitem{DeMarco2014}
F.~DeMarco, J.~Xuan, D.~Le~Berre, and M.~Monperrus, ``Automatic repair of buggy
  if conditions and missing preconditions with smt,'' in \emph{Proceedings of
  the 6th International Workshop on Constraints in Software Testing,
  Verification, and Analysis}, 2014.

\bibitem{Kim2013:PAR}
D.~Kim, J.~Nam, J.~Song, and S.~Kim, ``Automatic patch generation learned from
  human-written patches,'' in \emph{IEEE/ACM International Conference on
  Software Engineering (to appear)}, 2013.

\bibitem{LeGoues12:gp}
C.~Le~Goues, T.~Nguyen, S.~Forrest, and W.~Weimer, ``Genprog: A generic method
  for automatic software repair,'' \emph{IEEE Trans. Softw. Eng.}, vol.~38,
  no.~1, January 2012.

\bibitem{Nguyen13}
H.~D.~T. {Nguyen}, D.~{Qi}, A.~{Roychoudhury}, and S.~{Chandra}, ``Semfix:
  Program repair via semantic analysis,'' in \emph{2013 35th International
  Conference on Software Engineering (ICSE)}, 2013, pp. 772--781.

\bibitem{Meng11}
\BIBentryALTinterwordspacing
N.~Meng, M.~Kim, and K.~S. McKinley, ``Systematic editing: Generating program
  transformations from an example,'' in \emph{Proceedings of the 32nd ACM
  SIGPLAN Conference on Programming Language Design and Implementation}, ser.
  PLDI '11.\hskip 1em plus 0.5em minus 0.4em\relax New York, NY, USA:
  Association for Computing Machinery, 2011, pp. 329--342. [Online]. Available:
  \url{https://doi.org/10.1145/1993498.1993537}
\BIBentrySTDinterwordspacing

\bibitem{Meng13}
N.~Meng, M.~Kim, and K.~McKinley, ``Lase: Locating and applying systematic
  edits by learning from examples,'' in \emph{Proceedings of the 2013
  International Conference on Software Engineering}, ser. ICSE '13.\hskip 1em
  plus 0.5em minus 0.4em\relax IEEE Press, 2013, p. 502?511.

\bibitem{Bader19}
\BIBentryALTinterwordspacing
J.~Bader, A.~Scott, M.~Pradel, and S.~Chandra, ``Getafix: Learning to fix bugs
  automatically,'' \emph{Proc. ACM Program. Lang.}, vol.~3, no. OOPSLA, Oct.
  2019. [Online]. Available: \url{https://doi.org/10.1145/3360585}
\BIBentrySTDinterwordspacing

\bibitem{Tufano19}
\BIBentryALTinterwordspacing
M.~Tufano, C.~Watson, G.~Bavota, M.~D. Penta, M.~White, and D.~Poshyvanyk, ``An
  empirical study on learning bug-fixing patches in the wild via neural machine
  translation,'' \emph{ACM Trans. Softw. Eng. Methodol.}, vol.~28, no.~4, Sep.
  2019. [Online]. Available: \url{https://doi.org/10.1145/3340544}
\BIBentrySTDinterwordspacing

\bibitem{Mechtaev16}
S.~{Mechtaev}, J.~{Yi}, and A.~{Roychoudhury}, ``Angelix: Scalable multiline
  program patch synthesis via symbolic analysis,'' in \emph{2016 IEEE/ACM 38th
  International Conference on Software Engineering (ICSE)}, 2016, pp. 691--701.

\bibitem{Mechtaev18}
\BIBentryALTinterwordspacing
S.~Mechtaev, M.-D. Nguyen, Y.~Noller, L.~Grunske, and A.~Roychoudhury,
  ``Semantic program repair using a reference implementation,'' in
  \emph{Proceedings of the 40th International Conference on Software
  Engineering}, ser. ICSE '18.\hskip 1em plus 0.5em minus 0.4em\relax New York,
  NY, USA: Association for Computing Machinery, 2018, pp. 129--139. [Online].
  Available: \url{https://doi.org/10.1145/3180155.3180247}
\BIBentrySTDinterwordspacing

\bibitem{oj}
``{Online Judge},'' \url{https://onlinejudge.org}, 2020.

\bibitem{leetcode}
``{LeetCode},'' \url{https://leetcode.com}, 2020.

\bibitem{Isong01}
J.~Isong, ``Developing an automated program checkers,'' \emph{J. Comput. Sci.
  Coll.}, vol.~16, no.~3, pp. 218--224, Mar. 2001.

\bibitem{Reek96}
\BIBentryALTinterwordspacing
K.~A. Reek, ``A software infrastructure to support introductory computer
  science courses,'' \emph{SIGCSE Bull.}, vol.~28, no.~1, pp. 125--129, Mar.
  1996. [Online]. Available: \url{https://doi.org/10.1145/236462.236524}
\BIBentrySTDinterwordspacing

\bibitem{Jackson97}
\BIBentryALTinterwordspacing
D.~Jackson and M.~Usher, ``Grading student programs using assyst,''
  \emph{SIGCSE Bull.}, vol.~29, no.~1, pp. 335--339, Mar. 1997. [Online].
  Available: \url{https://doi.org/10.1145/268085.268210}
\BIBentrySTDinterwordspacing

\bibitem{Luck99}
M.~Luck and M.~Joy, ``A secure on-line submission system,'' \emph{Software:
  Practice and Experience}, vol.~29, no.~8, pp. 721--740, 1999.

\bibitem{Edwards03}
\BIBentryALTinterwordspacing
S.~H. Edwards, ``Improving student performance by evaluating how well students
  test their own programs,'' \emph{J. Educ. Resour. Comput.}, vol.~3, no.~3,
  pp. 1--es, Sep. 2003. [Online]. Available:
  \url{https://doi.org/10.1145/1029994.1029995}
\BIBentrySTDinterwordspacing

\bibitem{Adam80}
\BIBentryALTinterwordspacing
A.~Adam and J.-P. Laurent, ``Laura, a system to debug student programs,''
  \emph{Artificial Intelligence}, vol.~15, no.~1, pp. 75--122, 1980. [Online].
  Available:
  \url{http://www.sciencedirect.com/science/article/pii/0004370280900235}
\BIBentrySTDinterwordspacing

\bibitem{Tillmann13}
N.~{Tillmann}, J.~{de Halleux}, T.~{Xie}, S.~{Gulwani}, and J.~{Bishop},
  ``Teaching and learning programming and software engineering via interactive
  gaming,'' in \emph{2013 35th International Conference on Software Engineering
  (ICSE)}, 2013, pp. 1117--1126.

\bibitem{Tillmann14}
\BIBentryALTinterwordspacing
N.~Tillmann, J.~Bishop, N.~Horspool, D.~Perelman, and T.~Xie, ``Code hunt:
  Searching for secret code for fun,'' in \emph{Proceedings of the 7th
  International Workshop on Search-Based Software Testing}, ser. SBST
  2014.\hskip 1em plus 0.5em minus 0.4em\relax New York, NY, USA: Association
  for Computing Machinery, 2014, pp. 23--26. [Online]. Available:
  \url{https://doi.org/10.1145/2593833.2593838}
\BIBentrySTDinterwordspacing

\bibitem{Zimmerman15}
K.~Zimmerman and C.~R. Rupakheti, ``An automated framework for recommending
  program elements to novices (n),'' in \emph{2015 30th IEEE/ACM International
  Conference on Automated Software Engineering (ASE)}, 2015, pp. 283--288.

\bibitem{Kim16}
\BIBentryALTinterwordspacing
D.~Kim, Y.~Kwon, P.~Liu, I.~L. Kim, D.~M. Perry, X.~Zhang, and
  G.~Rodriguez-Rivera, ``Apex: Automatic programming assignment error
  explanation,'' \emph{SIGPLAN Not.}, vol.~51, no.~10, pp. 311--327, Oct. 2016.
  [Online]. Available: \url{https://doi.org/10.1145/3022671.2984031}
\BIBentrySTDinterwordspacing

\bibitem{Wang18}
\BIBentryALTinterwordspacing
K.~Wang, R.~Singh, and Z.~Su, ``Search, align, and repair: Data-driven feedback
  generation for introductory programming exercises,'' \emph{SIGPLAN Not.},
  vol.~53, no.~4, pp. 481--495, Jun. 2018. [Online]. Available:
  \url{https://doi.org/10.1145/3296979.3192384}
\BIBentrySTDinterwordspacing

\bibitem{Song19}
\BIBentryALTinterwordspacing
D.~Song, M.~Lee, and H.~Oh, ``Automatic and scalable detection of logical
  errors in functional programming assignments,'' \emph{Proc. ACM Program.
  Lang.}, vol.~3, no. OOPSLA, Oct. 2019. [Online]. Available:
  \url{https://doi.org/10.1145/3360614}
\BIBentrySTDinterwordspacing

\bibitem{Hu19}
Y.~Hu, U.~Z. Ahmed, S.~Mechtaev, B.~Leong, and A.~Roychoudhury, ``Re-factoring
  based program repair applied to programming assignments,'' in \emph{2019 34th
  IEEE/ACM International Conference on Automated Software Engineering (ASE)},
  2019, pp. 388--398.

\bibitem{Birch:2019aa}
\BIBentryALTinterwordspacing
G.~Birch, B.~Fischer, and M.~Poppleton, ``Fast test suite-driven model-based
  fault localization with application to pinpointing defects in student
  programs,'' \emph{Software \& Systems Modeling}, vol.~18, no.~1, pp.
  445--471, 2019. [Online]. Available:
  \url{https://doi.org/10.1007/s10270-017-0612-y}
\BIBentrySTDinterwordspacing

\bibitem{Yasunaga20}
\BIBentryALTinterwordspacing
M.~Yasunaga and P.~Liang, ``Graph-based, self-supervised program repair from
  diagnostic feedback,'' in \emph{Proceedings of the 37th International
  Conference on Machine Learning}, ser. Proceedings of Machine Learning
  Research, vol. 119.\hskip 1em plus 0.5em minus 0.4em\relax PMLR, 13--18 Jul
  2020, pp. 10\,799--10\,808. [Online]. Available:
  \url{http://proceedings.mlr.press/v119/yasunaga20a.html}
\BIBentrySTDinterwordspacing

\bibitem{Gupta17}
R.~Gupta, S.~Pal, A.~Kanade, and S.~Shevade, ``Deepfix: Fixing common c
  language errors by deep learning,'' in \emph{Proceedings of the Thirty-First
  AAAI Conference on Artificial Intelligence}, ser. AAAI'17.\hskip 1em plus
  0.5em minus 0.4em\relax AAAI Press, 2017, pp. 1345--1351.

\end{thebibliography}
